\documentclass[11pt]{article}
\RequirePackage{graphicx}
\RequirePackage{graphicx}
\RequirePackage{graphicx}
\RequirePackage{mathptmx}      % use Times fonts if available on your TeX system
\RequirePackage{flushend}
\RequirePackage[numbers,sort&compress]{natbib}
\RequirePackage[colorlinks,citecolor=blue,urlcolor=blue,linkcolor=blue]{hyperref}
\RequirePackage[T1]{fontenc}
\newcommand {\sla}[1]{ #1 \!\!\!\!/}
\def\ds{\displaystyle}

\begin{document}

\title{$\Delta(1232)$  Resonance Contribution to Two-Photon Exchange in Electron-Proton Scattering Revisited}

\author{
Hai-Qing Zhou$^1$ \footnote{zhouhq@seu.edu.cn}, and Shin Nan Yang$^2$\footnote{snyang@phys.ntu.edu.tw} \\
$^1$Department of Physics,
Southeast University, NanJing 211189, China\\
$^2$Department of Physics and Center for Theoretical Sciences,\\
National Taiwan University, Taipei 10617, Taiwan }
\date{\today}

\maketitle

\begin{abstract}
We revisit the question of the contributions of two-photon exchange
with $\Delta(1232)$  excitation to the electron-proton scattering in
a hadronic model. Three improvements over the previous calculations
are made, namely,  correct vertex function for $\gamma
N\rightarrow\Delta$, realistic $\gamma N\Delta$  form factors, and
coupling constants. The discrepancy between the values of $R\equiv
\mu_p G_E/G_M$ extracted from Rosenbluth technique and polarization
transfer method can be reasonably accounted for if the data of
Andivahis {\it et al.} (Phys. Rev.
D 50, 5491 (1994)) are analyzed. However,
substantial discrepancy remains if the data of Qattan {\it et al.}
(nucl-ex/0610006) are used. For the ratio $R^\pm$ between $e^\pm p$
scatterings, our predictions appear to be in satisfactory agreement
with the preliminary data from VEPP-3. The agreement between our
model predictions and the recent  measurements on single spin
asymmetry, transverse and longitudinal recoil proton polarizations
ranges from good to poor.
\end{abstract}

\section{Introduction}
\label{intro}

Proton is the only stable hadron and hence most amenable to
experimental measurement in the hadron structure study.
Determination of the proton form factors via electron elastic
scattering started in the 1950's. Nearly half of a century of
efforts yield the so-called {\it scaling law}, i.e., $R=\mu_pG_E/G_M
\sim 1$ for $Q^2<6 $ GeV$^2$, where $\mu_p$, $G_E$, and $G_M$ are
the magnetic moment, Sach's electric and magnetic form factors of
the proton, respectively, as often quoted in textbooks. The
measurements leading to the scaling law were all obtained from
analyses of the data based on the one-photon exchange (OPE)
approximation.

In the OPE approximation, the proton's electric
  and magnetic   form factors (FFs) can be extracted
from the reduced differential cross section $\sigma_R$ of the
electron-proton $\it (ep)$ elastic scattering as one has
\begin{eqnarray}
 \sigma_R(Q^2,\epsilon) \equiv\frac
{d\sigma}{d\Omega_{lab}}\frac{\epsilon(1+\tau)}{\tau\sigma_{Mott}} =
G_M^2 + \frac{\epsilon}{\tau}G_E^2, \label{diffCr}
\end{eqnarray}
where $\tau=Q^2/4M_N^2,\,\,
\epsilon^{-1}=1+2(1+\tau)tan^2\theta/2,\,\, Q^2=-q^2$  the momentum
transfer squared, $M_N$ the nucleon mass, $\theta$   the laboratory
scattering angle, $0\le\epsilon\le 1$, and $\sigma_{Mott}$ is the
Mott cross section for the scattering from a point particle,
\begin{eqnarray}
\sigma_{Mott}\equiv\frac{\alpha^2E_3 cos^2\frac{\theta}{2}}{4E_1^3
sin^4\frac{\theta}{2}},
\end{eqnarray}
with $E_1$ and $E_3$ the initial and final electron energies and
$\alpha=e^2/4\pi$ the electromagnetic fine structure constant. For
fixed $Q^2$, varying angle $\theta$, i.e. $\epsilon$, and adjusting
incoming electron energy as needed to plot $\sigma_R$ versus
$\epsilon$ will give the FFs, a method often called the Rosenbluth,
or longitudinal-transverse (LT), separation technique.

The good times with scaling law ended when, at the turn of this
century, a polarization transfer (PT) experiment carried out at JLab
yielded values of $R$  markedly different from 1 in the  range of
$0.2 < Q^2< 8.5 $ GeV $^2$ \cite{Jones00,Gayou02,Puckett10,Zhan11,Ron11}. The polarization
experiment is based on a result shown in \cite{Akhiezer58,Akhiezer74} that,
again in the OPE approximation, the ratio $R$ can be accessed in
$ep$ scattering with longitudinally polarized electron  by measuring
the polarizations of the recoiled proton parallel $P_l$ and
perpendicular $P_t$ to the proton momentum in the scattering plane,
\begin{eqnarray}
R=\frac{\mu_p
G_E}{G_M}=-\mu_p\sqrt{\frac{\tau(1+\epsilon)}{2\epsilon}}\frac{P_t}{P_l}.\label{polR}
\end{eqnarray}
Polarization transfer experiment of this kind is only possible
recently at JLab. It came as a big surprise that the  PT experiments
yield values of $R$ deviate substantially from 1.  It prompts
intensive efforts, both experimentally and theoretically. The
readers are referred to recent reviews
\cite{Carlson07,Arrington11,Yang13} for details on these
developments. In addition, a comprehensive exposition of the application of the soft-collinear effective theory
(SCET) to the study of the two-photon exchange (TPE) corrections to the electron-proton scattering in the region where the kinematical variables describing the elastic ep scattering are moderately large momentum scales relative to the soft hadronic scale is presented in \cite{Kivel13}.

On the experimental side, a new global analysis of the world's cross
section data was carried out in \cite{Arrington03}. It is found that
the great majority of the measured cross sections were consistent
with each other and the disagreement with polarization transfer
measurements remains. A set of extremely high precision measurements
of $R$ was later performed using a modified Rosenbluth technique
\cite{Qattan05,Qattan06}, with the detection of recoil proton to minimize the
systematic uncertainties, and the discrepancy is again confirmed.

The immediate step taken, on the theoretical side, was to carefully
reexamine the radiative corrections which were known to be as large
as $30\%$ of the uncorrected cross section in certain kinematics. Of
various radiative corrections, only proton-vertex and two-photon
exchange (TPE) corrections contained $\epsilon$ dependence. The
proton-vertex corrections had been investigated thoroughly in
\cite{MT00} and found to be negligible. Realistic evaluations of the
TPE corrections are hence called for to see whether they can explain
the discrepancy.

A semi-quantittative analysis \cite{Guichon03} quickly established
that the discrepancy can possibly be explained by a two-photon
exchange correction which would not destroy the linearity of the
Rosenbluth plot. The ensuing theoretical investigation of the
two-photon exchange effects include hadronic
\cite{Blunden03,Kondra05,Blunden05} and partonic model
\cite{Chen04,Afana05} calculations, phenomenological
parametrizations \cite{Chen07,BK07}, dispersion approach
\cite{BK06,BK08,BK11,BK12,BK14}, and pQCD calculations \cite{BK09,Kivel09}. They all
have found that TPE effects can account for more than half of the
discrepancy.

The hadronic model calculations of the effects of TPE with  nucleon
intermediate states, denoted as TPE-N hereafter, have established
that it is important to employ realistic $\gamma NN$ form factors
\cite{Blunden03,Blunden05}. For the inelastic contributions, it has been
demonstrated in \cite{BK05} that $\Delta(1232)$ dominates in the
case of target-normal spin asymmetry. The effects of TPE  with
$\Delta$ excitation, denoted as TPE-$\Delta$ hereafter, in the cross
sections and the form factors have been studied  in
\cite{Kondra05,BK12,BK14}. However, there are rooms for improvement in
three aspects of these calculations to arrive at a reliable estimate
of the TPE-$\Delta$ effects. First, as was pointed out in
\cite{Zhou10}, the expression for the vertex function of $\gamma N
\rightarrow \Delta$ used in \cite{Kondra05} has the incorrect sign
for the Coulomb quadrupole coupling, though it was not considered
in \cite{BK12}. Next is that the $\gamma N\Delta$ form factors
employed  in  \cite{Kondra05} are not realistic which, as we
learn in the case of TPE-N, needs to be studied. Lastly, both
\cite{Kondra05,BK12} set the Coulomb quadrupole coupling  to be
zero, which is again not satisfactory since  recent pion
electroproduction experiments and the LQCD results indicate that the
ratio of Coulomb quadrupole (C2) over magnetic dipole (M1), denote
by $R_{SM}=C2/M1$ grows more negative with increasing $Q^2$
\cite{Joo02,Sparveris05,Ungaro06,Pascalu07,Alexandrou05}. The theoretical understandings
of the discrepancy between LT and PT experiments, as well as the TPE
contributions are still ongoing. It is important to have the results
from various model calculations as accurate as possible so as to
understand the strength and the weakness of different approaches and
shed light for the further study. Accordingly, we set out in this
study to improve the previous calculations of the effects of
TPE-$\Delta$ excitation  \cite{Kondra05} on the three aspects
described in the above.

This article is organized as follows. In Sec. II, we give the
explicit expression for the amplitude of two-photon exchange with
$\Delta$ in the intermediate states and elaborate on the details of
the three improvements we will implement. They are, (1) the correct
expression for the $\gamma N\rightarrow\Delta$ with Coulomb
quardrupole coupling; (2) the realistic $\gamma N\rightarrow\Delta$
form factors; and (3) Coulomb quadrupole $\gamma
N\rightarrow\Delta$ coupling constant as given by the recent
experiment. Results with the implementation of each of these three
improvements are presented in Sec. III and compared with those
obtained in \cite{Kondra05} to demonstrate their importance. We
then proceed to present results, obtained with all three
improvements combined, for reduced cross sections, extracted $R$ in
LT method, ratio $R^\pm$ between positron-proton and electron-proton
cross sections, single spin asymmetries, longitudinal and transverse
polarizations of the recoil proton $P_l, P_t$ and their ratio
$R_{PT}=-\mu_p \sqrt{\tau(1+\epsilon)/2\epsilon}P_t/P_l$. In
Sec. IV, we summarize our results.

\section{Two-photon exchange with $\Delta(1232)$ excitation in elastic electron-proton scattering}
\begin{figure}
\center{
\includegraphics[width=0.45\textwidth]{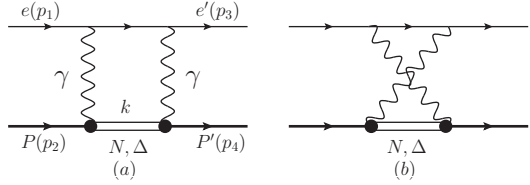}
\caption{Two-photon exchange  diagrams with $\Delta$ excitation for elastic
{\it ep} scattering.} \label{TPE-Delta}
}
\end{figure}

In this section, we discuss the evaluation of the two-photon
exchange (TPE) diagrams with $\Delta(1232)$ excitation TPE-$\Delta$,
as depicted in Fig. ~\ref{TPE-Delta},
 in a
simple hadronic model. The amplitude for the box diagram in Fig.~\ref{TPE-Delta}(a) is given as,

\begin{eqnarray}
M^{(a,\Delta)}&=&-i\int\frac{d^4k}{(2\pi)^4}\overline{u}(p_3)(-ie\gamma_{\mu})\frac{i(\sla{p}_1+\sla{p}_2-\sla{k})}
{(p_1+p_2-k)^2-m_e^2+i\varepsilon}  \nonumber \\ & &
\times(-ie\gamma_{\nu})u(p_1)\frac{-i}{(p_4-k)^2+i\varepsilon}
\times\frac{-i}{(k-p_2)^2+i\varepsilon}  \nonumber \\& &
\times\overline{u}(p_4)\Gamma^{\mu\alpha}_{\gamma\Delta\rightarrow N}(k,p_{4}-k)\frac{-i(\sla{k}+M_{\Delta})P_{\alpha\beta}^{3/2}(k)}
{k^2-M_{\Delta}^2+i\varepsilon} \nonumber \\& &
\times \Gamma^{\nu\beta}_{\gamma N\rightarrow \Delta}(k,k-p_{2})u(p_2),\label{eq:box-amp}
\end{eqnarray}
where
\begin{eqnarray}
P_{\alpha\beta}^{3/2}(k)=g_{\alpha\beta}-\frac{\gamma_{\alpha}\gamma_{\beta}}{3}
-\frac{(\sla{k}\gamma_{\alpha}k_{\beta}+k_{\alpha}\gamma_{\beta}\sla{k})}{3k^2},
\end{eqnarray}
is the spin-3/2 projector. Amplitude for the cross-box diagram Fig.~\ref{TPE-Delta}(b) can be written down in similar manner. The
amplitude in Eq. (\ref{eq:box-amp}) is IR finite because when the
four-momentum of the photon approaches zero, the $\gamma N\Delta$
vertex  functions $\Gamma's$ also approaches zero. Therefore we do
not have to include an infinitesimal photon mass in the photon
propagators to regulate the IR divergence in Eq. (\ref{eq:box-amp}).
The vertex functions $\Gamma's$ for $\gamma\Delta\rightarrow N$ and
$\gamma N\rightarrow\Delta$ are defined by
\begin{eqnarray}
 \overline{u}(p+q)\Gamma^{\mu\alpha}_{\gamma\Delta\rightarrow
N}(p,q)u^\Delta_{\alpha}(p)&=&-ie\langle
N(p+q)|J^\mu_{EM}|\Delta(p)\rangle, \label{Delta-to-N} \\
 \overline{u}^\Delta_{\beta}(p)\Gamma^{\nu\beta}_{\gamma
 N\rightarrow
\Delta}(p,q)u(p-q)&=&-ie\langle \Delta(p)|J^\nu_{EM}|N(p-q)\rangle,
\label{N-to-Delta}
\end{eqnarray}
where the $q's$ in both $\Gamma^{\mu\alpha}_{\gamma\Delta\rightarrow
N}(p,q)$ and $\Gamma^{\beta\nu}_{\gamma N\rightarrow \Delta}$ refer
to the {\it incoming} momentum of the photon, as in \cite{Kondra05}.

We now elaborate, in the followings, on the three improvements over
the previous calculations we will carry out in this study.

\subsection{Relation between vertex functions  of $ \gamma\Delta\rightarrow  N $ and
$\gamma N\rightarrow \Delta$}

The correct relations between the two vertex functions for
$\gamma\Delta\rightarrow N$ and $\gamma N\rightarrow \Delta$ are
\begin{equation}
\Gamma_{\gamma\Delta\rightarrow N}(p,q)=-\gamma_{0}[\Gamma_{\gamma
 N\rightarrow\Delta}(p,-q)]^{\dagger}\gamma_{0}, \label{ours}
\end{equation}
with $q's$ in both sides of the above Eq. (\ref{ours}) denote the
{\it incoming} momentum of the photon. It follows from  the fact
that electromagnetic current is Hermitian. However, in
\cite{Kondra05,Tjon09} the following relation between
$\Gamma_{\gamma N\rightarrow\Delta}$ and
$\Gamma_{\gamma\Delta\rightarrow N}$ has been used:
\begin{equation}
\Gamma_{\gamma\Delta\rightarrow N}(p,q)=\gamma_{0}[\Gamma_{\gamma
 N\rightarrow\Delta}(p,q)]^{\dagger}\gamma_{0}. \label{theirs}
\end{equation}

Specifically, with the inclusion of the form factors, vertex
function $\Gamma_{\gamma \Delta \rightarrow N}^{\mu \alpha}$ takes
the form
\protect\footnotemark[1] \protect\footnotetext[1]{In our
definition, there is a global minus sign difference with that used
in \cite{Kondra05}, since such global minus will not change the
results, such global minus sign in the choice of $g_i$ of
\cite{Kondra05} is ignored. }
\begin{eqnarray}
\Gamma_{\gamma \Delta \rightarrow N}^{\mu \alpha}(p,q)
 &=& -i \sqrt{\frac{2}{3}}{\ds \frac{e}{2 M_{\Delta}^2}} \bigg\{
g_1 F^{(1)}_\Delta(q^2) [ g^{\mu \alpha} \sla{p} \sla{q} - \nonumber \\ & &
p^\mu \gamma^\alpha \sla{q}
- \gamma^\mu \gamma^\alpha p \cdot q + \gamma^\mu \sla{p} q^\alpha ] \nonumber\\ & &+ g_2 F^{(2)}_\Delta(q^2)\left[\, p^\mu q^\alpha - g^{\mu \alpha} p \cdot q\, \right] \nonumber\\&&
+ (g_3/M_{\Delta})F^{(3)}_\Delta(q^2) [ q^2 (p^\mu
\gamma^\alpha - g^{\mu \alpha} \sla{p}) \nonumber \\ & &
 + q^\mu (q^\alpha \sla{p} -
\gamma^\alpha p \cdot q ) ] \bigg\} \gamma_5.
\label{Vertex-D3-D-to-N}
\end{eqnarray}
Eq. (\ref{ours}) then leads to
\begin{eqnarray}
\Gamma_{\gamma N \rightarrow \Delta}^{\nu \beta}(p,q)
&=& -i\sqrt{\frac{2}{3}} {\ds \frac{e}{2 M_{\Delta}^2}} \gamma_5\bigg\{
g_1F^{(1)}_\Delta(q^2) [ g^{\nu \beta} \sla{q}\sla{p} \nonumber\\&&
-p^\nu \sla{q}\gamma^\beta
- \gamma^\beta \gamma^\nu  p \cdot q + \sla{p}\gamma^\nu  q^\beta ] \nonumber\\ &&
+ g_2F^{(2)}_\Delta(q^2) [ p^\nu q^\beta - g^{\nu \beta} p \cdot q ] \nonumber\\ &&
- (g_3/M_{\Delta})F^{(3)}_\Delta(q^2) [ q^2 (p^\nu
\gamma^\beta - g^{\nu \beta} \sla{p})  \nonumber\\ &&
+ q^\nu (q^\beta \sla{p} -
\gamma^\beta p \cdot q ) ] \bigg\}, \label{Vertex-D3-N-to-D}
\end{eqnarray}
where at $Q^2=0$, $g_i's$ are related to the conventionally used
magnetic dipole $G^*_M$, electric quadrupole $G^*_E$, and Coulomb
quardrupole couplings $G^*_C$ form factors by \cite{Pascalu07},

\begin{eqnarray}
g_1 &=& \frac{3M_\Delta^2}{M_N(M_\Delta+M_N)}(G^*_M-G^*_E)  \nonumber\\
g_2 &=& \frac{3M_\Delta^2(M_\Delta+3M_N)}{M_N(M_\Delta^2-M_N^2)}G^*_E+\frac{3M_\Delta^2}{M_N(M_\Delta+M_N)}G^*_M \nonumber\\
g_3 &=&
-\frac{3M_\Delta^2}{M_N(M_\Delta+M_N)}\left(-\frac{M_\Delta+M_N}{(M_\Delta-M_N)}G^*_C+\frac{4M_\Delta^2}{(M_\Delta-M_N)^2}G^*_E\right
),  \nonumber\\ &&
\label{relation-gG}
\end{eqnarray}

However, if Eq. (\ref{theirs}) is used, then one would get an
expression for $\Gamma_{\gamma N  \rightarrow \Delta}$ which would
lead to the last term in Eq. (\ref{Vertex-D3-N-to-D}) to carry a
different sign, namely, the negative sign in front of $g_3$ in Eq.
(\ref{Vertex-D3-N-to-D}) becomes positive. Since in both
\cite{Kondra05,Tjon09}   $g_3$ was set to zero,  this sign problem
would not affect the results presented therein.

The difference between Eq. (\ref{ours}) and Eq. (\ref{theirs})
incurs significant discrepancy in the results, in the case of
corrections of $\gamma Z$ exchange with $\Delta$ excitation to the
parity-violating electron-proton scattering, obtained in
\cite{Nagata09,Zhou10} and \cite{Tjon09} at the forward angles and
higher $Q^2$. Similar situation can be expected to arise in the
parity-conserving $ep$ scattering as well. In this study we use Eq.
(\ref{ours}) because it is derived from the fact that the currents
are Hermitian.

\subsection{Realistic form factors for $\gamma N\Delta$ vertex}
As demonstrated in \cite{Blunden03,Blunden05}, the estimated contribution of TPE-N is reliable only if the employed nucleon form factors are realistic, similar situation can be
expected to arise in the case with $\Delta$ intermediate states.

In \cite{Kondra05}, all three form factors $(F_\Delta^{(i)},\,\,
i=1,3)$ in Eqs. (\ref{Vertex-D3-D-to-N}, \ref{Vertex-D3-N-to-D}) are
assumed to take the same form as
\begin{equation}
F^{(i)}_\Delta(q^2) = F(q^2)= \left
(\frac{-\Lambda^2_1}{q^2-\Lambda^2_1}\right )^2, \,
(i=1,3),\label{D1}
\end{equation}
with $\Lambda_{1}$ = 0.84 GeV.

\begin{figure}
\center{
\includegraphics[width=0.45\textwidth]{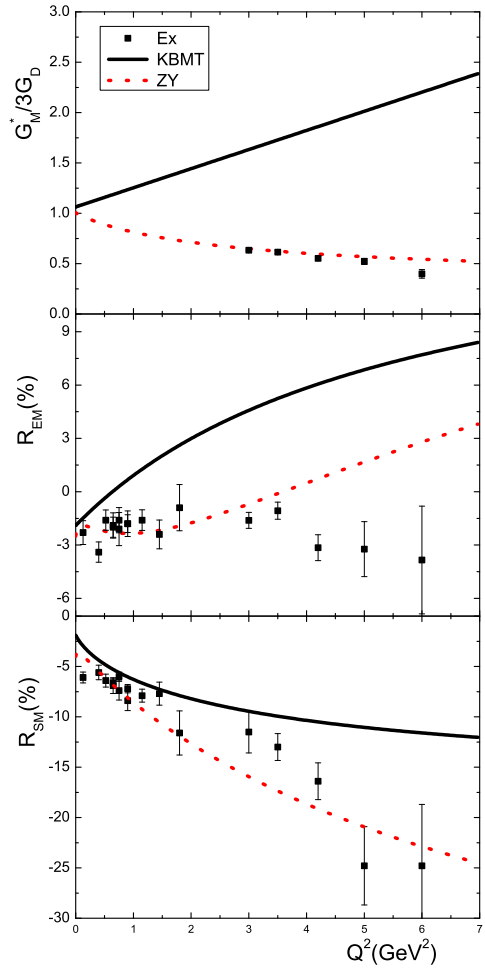}
\caption{
Comparison of the $\Delta$ form factor $G^*_M$, $R_{EM}$, and
$R_{SM}$ used in \cite{Kondra05} and this study with the
experimental data \cite{Joo02,Sparveris05,Ungaro06}.} \label{FFs-Delta}
}
\end{figure}

In this investigation,  the $\Delta$ form factors are taken to have
the following forms,
\begin{eqnarray}
F^{(1)}_{\Delta}&=&F^{(2)}_{\Delta}=\left(\frac{-\Lambda_{1}^2}{q^2-\Lambda_{1}^{2}}\right)^{2}
\frac{-\Lambda_{3}^2}{q^2-\Lambda_{3}^{2}},\,\nonumber\\
F^{(3)}_{\Delta}&=&\left(\frac{-\Lambda_{1}^2}{q^2-\Lambda_{1}^{2}}\right)^{2}\frac{-\Lambda_{3}^2}{q^2-\Lambda_{3}^{2}}
\left [a \frac{-\Lambda_{2}^2}{q^2-\Lambda_{2}^{2}}+(1-a)
\frac{-\Lambda_{4}^2}{q^2-\Lambda_{4}^{2}}\right ],\,\nonumber\\
\label{D3}
\end{eqnarray}
with $\Lambda_{1}= 0.84 \, $GeV,$\,\Lambda_{2} = 2\,\,$ GeV$,\,
\Lambda_{3} = \sqrt{2}\,\, $GeV, $\Lambda_{4}$ = $0.2$ GeV,
$a = -0.3.$ In Fig.~\ref{FFs-Delta}, we compare the conventional
magnetic dipole ($G^*_M$), the ratio of electric quadrupole (E2)
over magnetic dipole (M1),  and the ratio of Coulomb quadrupole
(C2) over magnetic dipole (M1), denoted by $R_{EM}$ and $R_{SM}$
\cite{Pascalu07}, respectively, resulting from the form factors
given used in \cite{Kondra05} and this study, as given in Eqs.
(\ref{D1}, \ref{D3}), with the experimental data taken from
\cite{Joo02,Sparveris05,Ungaro06}. The black solid curves, labeled as KBMT, denote the
predictions as would be obtained with Eq. (\ref{D1}) as employed in
\cite{Kondra05}. They deviate strongly from the experimental data,
especially for $G^*_M$ and $R_{EM}$. The red dashed curves, labeled
as ZY, correspond to predictions as would be obtained with Eq.
(\ref{D3}) and used in our study, agree well with the data except
for $R_{EM}$ at  $Q^2\sim 4-6 \, $ GeV$^2$ where we purposely impose
the prediction of PQCD to have $R_{EM}$ to approach one when $Q^2$
become infinity.

\subsection{$\gamma N\Delta$ coupling constants}
 The parameters used in this study are taken as
$(g_1, g_2, g_3)=(6.59, 9.08, 7.12)$ which are extracted from the
most recent experiments \cite{Pascalu07}. In  contrast,
\cite{Kondra05} use $(g_1, g_2, g_3)=(7, 9 , 0)$.  The biggest
difference lies with $g_3$ which corresponds to the Coulomb
quardrupole coupling. Our value for $g_3$ is extracted from the most
recent experiments and is quite large. For the finite $g_3$ case,
since the corrected $N\rightarrow\Delta$ vertex function as given in Eq. (\ref{Vertex-D3-N-to-D}) has a minus sign in front of $g_3$, while it would be positive if the prescription for this vertex function given in \cite{Kondra05} is followed, significant difference in the predictions can be expected.

\section{Results and discussions}

The loop integrals with $\Delta$ intermediate state are infrared
safe. We use computer package ``FeynCalc'' \cite{Feyncalc} and
``LoopTools'' \cite{Looptools} to carry out the calculations of
integrals of Eq. (\ref{eq:box-amp}).

In this section, we will first give the results of our calculation
with each of the  three improvements on the $\Delta$ contribution
 implemented separately, to demonstrate the importance of using
correct $\gamma N\Delta$ vertex function, realistic form factors and
coupling constants. Then we will proceed to   present our results
with all three improvements implemented together, as well as
employing realistic $\gamma NN$ form factors used in
\cite{Blunden05}, for the unpolarized cross sections, extracted
ratio $R=\mu G_E/G_M$, ratio $R^{\pm}$ between $e^+p$ and $e^-p$
scatterings, single spin asymmetries $B_n$ and $A_n$,  and
polarization observables $P_l, P_t$, and $R_{PT}$, and compare them
with results and the model predictions of \cite{Pasquini04}, as well
as the data.

\subsection{Separate effects of the three improvements: correct
$\gamma N\Delta$ vertex function, realistic $\gamma N\Delta$ form
factors, and coupling constants}

As in \cite{Kondra05},  the corrections of the TPE to the
unpolarized reduced cross section can be quantified as,
\begin{eqnarray}  \sigma_R &=&[G_M^2 +
\frac{\epsilon}{\tau}G_E^2](1+\bar\delta_N+\delta_\Delta)\nonumber\\
&=& [G_M^2 + \frac{\epsilon}{\tau}G_E^2](1+
\Delta_{un}),\label{def-delta}
\end{eqnarray} where $\bar\delta_N=\delta_N-\delta_{IR}(MT)$, with $\delta_{IR}(MT)$
 the well-known Mo and Tsai's radiative
corrections \cite{MT61,MT69} which are removed from data in typical
experimental analyses.  $\Delta_{un}= \bar\delta_N+\delta_\Delta$
with $\delta_N$($\delta_\Delta$) denotes the correction obtained
from the two-photon exchange diagrams with nucleons ($\Delta's$) in
the intermediate states, respectively, as depicted in
Fig.~\ref{TPE-Delta}.

If we denote the Born scattering amplitude as $M_B$ and the
two-photon exchange amplitudes with nucleon and $\Delta$
intermediate states as $M_N^{2\gamma}$ and $M_\Delta^{2\gamma}$,
then  to the first order in the electromagnetic coupling
$\alpha=e^2/4\pi$, $\delta_{N,\Delta}$ are given as,
\begin{equation}
\delta_{N,\Delta}=2\frac{Re(M_B^\dagger
M_{N,\Delta}^{2\gamma})}{\mid M_B\mid^2}. \label{delta-def}
\end{equation}
\noindent $\delta_N$ was well studied in \cite{Blunden03,Blunden05}.
For $\delta_{\Delta}$ in Eq.~(\ref{delta-def}), we note that it is
linear in $M_{\Delta}^{2\gamma}$.  Since $\gamma N\Delta$ vertex
appears twice in $M_\Delta^{2\gamma}$,  $\delta_\Delta$ can then be
expressed in a quadratic form in the $\gamma N\Delta$ coupling
constants $g_i's$,
\begin{eqnarray}
\delta_\Delta=\Sigma_{i,j=1}^3{C_{ij}g_ig_j}.\label{quadratic-rel}
\end{eqnarray}
The values of $C_{ij}$'s   vs. $\epsilon$ at $Q^2=3 \,$GeV$^2$,
 are presented in
Table \ref{Cij-Ours}, where only those with $i\leq j$ are given
because $C_{ij}= C_{ji}$. It is seen that all $C_{i3}$'s are one to
two orders smaller than the rest. We find that the values of
$C_{i3}$'s are very sensitive $w.r.t.$ the form factors in that they
would become comparable to the others if form factors of Eq.
(\ref{D1}) are used.

In \cite{Kondra05}, they chose to write $\delta_\Delta = C_{MM}g^2_M
+ C_{ME}g_M g_E + C_Eg^2_E+C_Cg^2_C + C_{EC}g_Eg_C +C _{MC}g_Mg_C$
instead, where $g_{M}=g_1,\, g_{E}=g_2-g_1,\, g_{C}=g_3$. Our
numbers would agree with those presented in Table I of
\cite{Kondra05} if their form factors of Eq. (\ref{D1}) are
employed, wherein $C_{MC,EC}$ are found to be less than $10^{-10}$.
In fact, both $C_{MC,EC}$ should be identically zero when the
incorrect relation between $\Gamma_{\gamma N\rightarrow\Delta}$ and
$\Gamma_{\gamma\Delta\rightarrow N}$ of Eq. (\ref{theirs}) is used
because  one would then have $C_{i3} = -C_{3i}, \,(i\neq 3)$.

\begin{table*}[hbtp]
\center{
\begin{tabular}
{|p{45pt}|p{45pt}|p{45pt}|p{45pt}|p{45pt}|p{45pt}|p{45pt}|}
\hline $\varepsilon$& 10$^{4}$ C$_{11}$& 10$^{4}$C$_{12}$& 10$^{4}$C$_{22}$& 10$^{6}$C$_{13}$& 10$^{6}$C$_{23}$& 10$^{6}$C$_{33}$ \\
\hline 0.1&-0.053&2.974&-1.015&-5.847&0.560&0.036 \\
\hline0.2&0.121&2.737&-1.048&-4.616&0.543&0.066 \\
\hline0.3&0.245&2.518&-1.054&-3.647&0.640&0.097 \\
\hline0.4&0.333&2.305&-1.036&-2.957&0.838&0.131 \\
\hline0.5&0.391&2.089&-0.991&-2.580&1.140&0.170 \\
\hline0.6&0.427&1.857&-0.918&-2.582&1.570&0.217 \\
\hline0.7&0.445&1.596&-0.809&-3.112&2.186&0.279 \\
\hline0.8&0.551&1.278&-0.647&-4.547&3.153&0.371 \\
\hline0.9&0.462&0.824&-0.376&-8.317&5.123&0.554 \\
\hline
\end{tabular}
\caption{$C_{ij}$ of Eq. (\ref{quadratic-rel}) at $Q^2 = 3$ GeV$^2$ obtained with   correct
vertex $\gamma N\Delta$ function, realistic  $\gamma N\Delta$ and
$\gamma NN$ form factors, and coupling constants.} \label{Cij-Ours}
}
\end{table*}

\begin{figure*}[hbtp]
\center{
\includegraphics[width=0.45\textwidth]{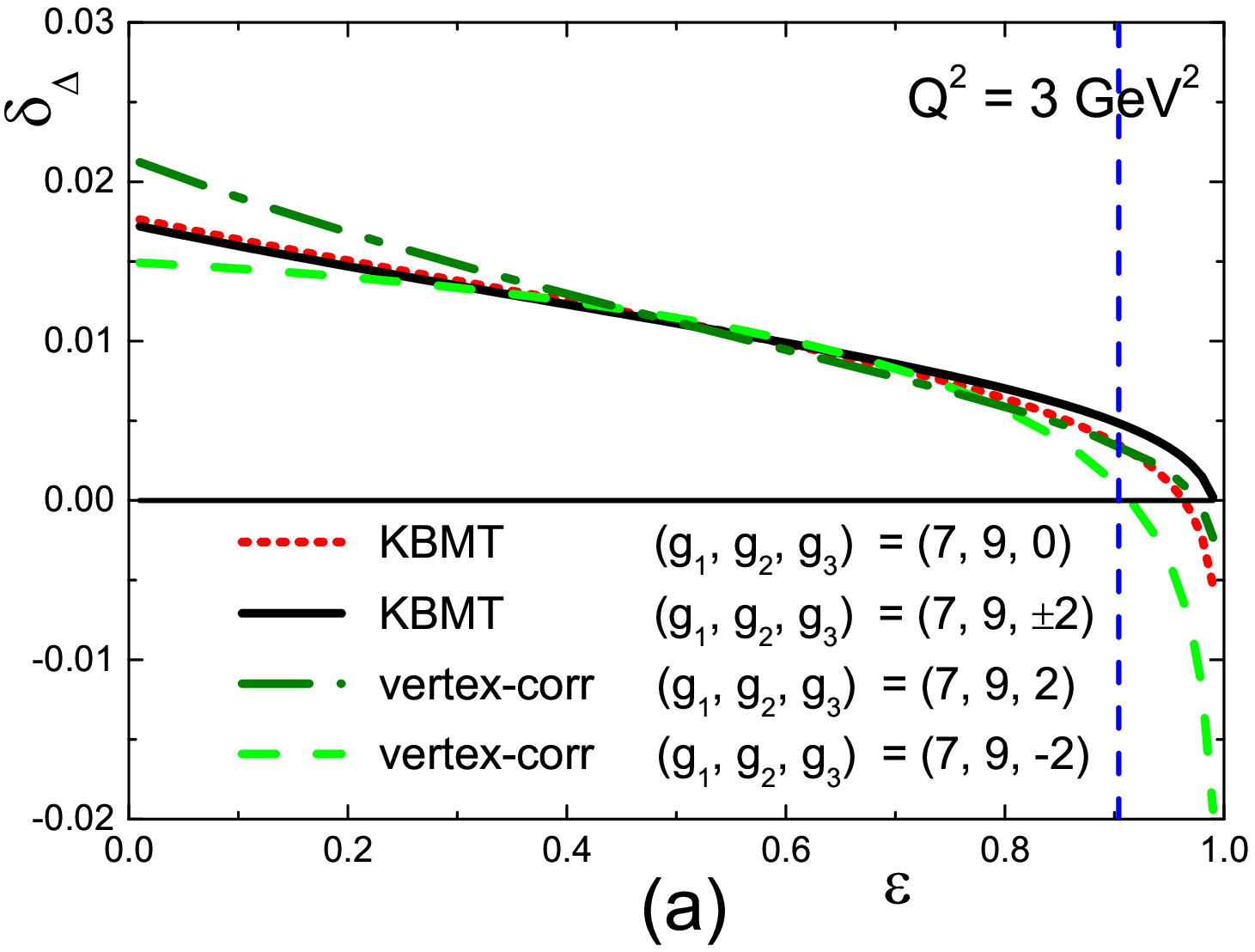}\includegraphics[width=0.45\textwidth]{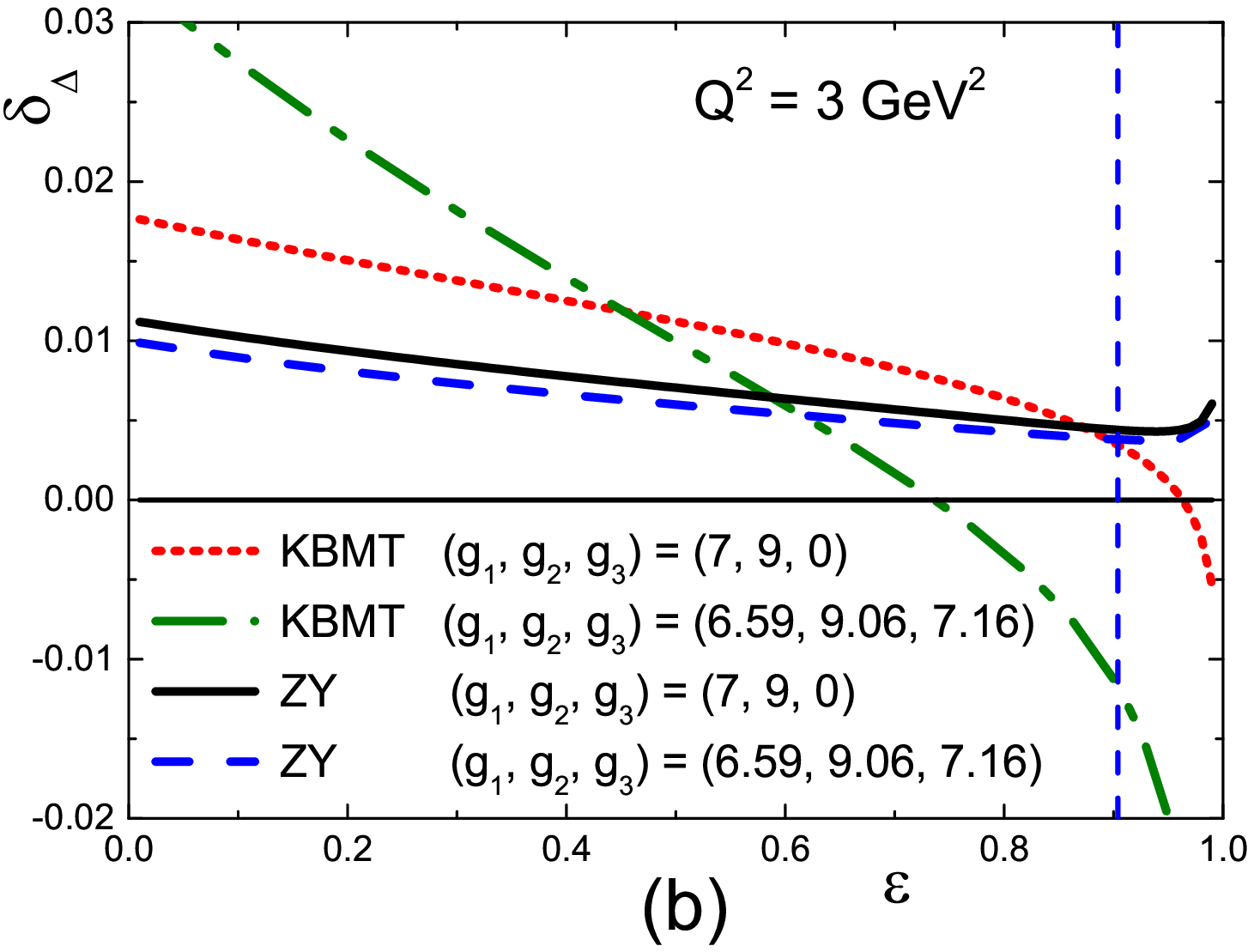}
\caption{$\delta_\Delta$ vs. $\epsilon$
at $Q^2 = 3 $ GeV$^2$. (a) With $\Delta$ form factors of Eq.
(\ref{D1}) and  coupling parameters $g_1 = 7,\, g_2 = 9$. The (red)
dotted and (black) solid curves correspond to $g_3 = 0$ and  $g_3 =
\pm2$, respectively, using     vertex relation of Eq. (\ref{theirs}).
(Green) dashed and (olive) dash-doted curves correspond to  $g_3=$
-2 and 2, obtained with   the correct vertex relation of Eq.
(\ref{ours}). (b) Dependence of $\delta_\Delta$ on $\epsilon$ with
the use of correct vertex function but different coupling constants
and form factors. The (red) dotted  and (olive) dash-doted curves,
labelled by  KBMT correspond to $g_1 = 7, g_2 = 9, g_3 = 0$ and $g_1
= 6.59, g_2 = 9.06, g_3 = 7.16$, respectively, both with the
$\Delta$ form factors of Eq. (\ref{D1}) employed in \cite{Kondra05}.
The (blue) dashed and (black) solid curves, labelled by ZY,
correspond to $g_1 = 7, g_2 = 9, g_3=0$ and $g_1 = 6.59, g_2 = 9.06,
g_3 = 7.16$ with the realistic $\Delta$ form factors of Eq.
(\ref{D3}).} \label{effects_from_vertex_paramerter_ff}
}
\end{figure*}

We first focus on the effects associated with the use of different
vertex functions given in Eqs. (\ref{ours}, \ref{theirs}). In Fig.~(\ref{effects_from_vertex_paramerter_ff}a), results for
$\delta_\Delta$ vs. $\epsilon$ at $Q^2 = 3\,\, $GeV$^2$,   with $g_1 =
7, g_2 = 9$, as considered in \cite{Kondra05}, are shown. The (red)
dotted  and the (black) solid curves, labeled as KBMT and using
their $\gamma N\Delta$ vertex relation Eq. (\ref{theirs}),
correspond to $g_3 = 0$ and $g_3 = \pm 2$, respectively.
   On the other hand,
  the (green) dashed and (olive) dash-doted curves, labeled as
  vertex-corr,
refer to $g_3=-2, 2$ using the correct vertex relation Eq.
(\ref{ours}). We see that even for small values of $|g_3| = 2$, it
is important to use the correct vertex function Eq.
(\ref{Vertex-D3-N-to-D}).

Fig.~(\ref{effects_from_vertex_paramerter_ff}b) illustrates the
 importance of employing realistic $\gamma N\Delta$ form factors and
 coupling constants, when the correct vertex functions are used.
The (red) dotted  and olive dash-doted curves, labeled by  KBMT,
 obtained with the $\Delta$ form factors Eq. (\ref{D1}) employed in
\cite{Kondra05}, correspond to $(g_1 = 7, g_2 = 9, g_3 = 0)$ and
$(g_1 = 6.59, g_2 = 9.06, g_3 = 7.16)$, respectively.  The set of
$(g_1 = 6.59, g_2 = 9.06, g_3 = 7.16)$ is the most recent one
extracted from experiments \cite{Pascalu07}. The difference between
the dotted and dashed curves arises solely from different values of
$g_3$ used. The (blue) dashed and (black) solid curves, labeled by
ZY and obtained with the realistic $\Delta$ form factors Eq.
(\ref{D3}), correspond to $(g_1 = 7, g_2 = 9, g_3=0)$ and $(g_1 =
6.59, g_2 = 9.06, g_3 = 7.16)$, respectively. The large differences
between   (red) dotted and (black) solid curves, and  (green)
dash-dotted and (blue) dashed curves, are attributed to the
different form factors used. However, one notes that the (black)
solid and (blue) dashed curves are very close to each other which
implies that once the realistic form factors are employed, the
effect of Coulomb quadrupole coupling is greatly reduced.

Hereafter, all the results to be given are obtained with the use of
correct $\gamma N\Delta$ vertex function, realistic form factors,
and coupling constants, unless otherwise specified.

\begin{figure}[hbtp]
\center{
\includegraphics[width=0.45\textwidth]{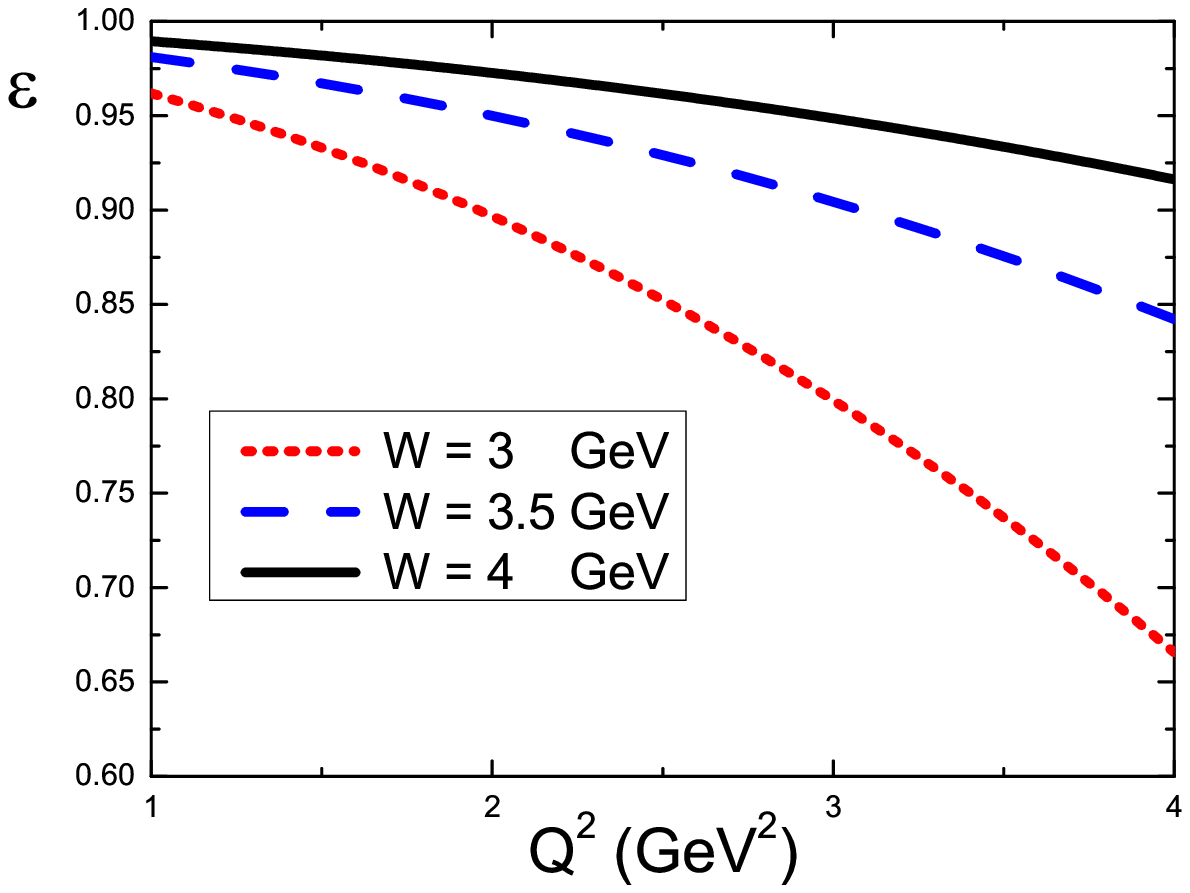}
\caption{$\epsilon$ vs. $Q^2$ for $W \sim 3 - 4$ GeV. }
 \label{range of epsilon}
}
\end{figure}

Recently, it has been assumed in \cite{BK08} that for $s=(p_1+p_2)^2 \rightarrow \infty$ (Regge limit), which leads to $\epsilon \rightarrow 1$, the TPE correction to $ep$ sccattering amplitude should vanish.
The assumption is made so that an {\it unsubtracted} fixed-$t$ dispersion relation can be written down for the TPE amplitude. Subsequently, such an assumption has been employed in various analyses \cite{BK11,Guttmann11,BK12} to extract TPE corrections from experimental data. Whether such a assumption is valid  remains to be substantiated.  The calculations of pQCD \cite{Kivel09,BK09} and SCET \cite{Kivel13} do support such an assumption. Nevertheless,
it is not clear whether their results would hold up in the soft hadronic scale. In fact,  the results of the GPD calculation, shown in Fig. 8 of \cite{Afana05} are not in line with such an assumption, \textcolor{black}{albeit the deviation is small}. Our results for TPE-N, which agree with those reported in \cite{Blunden03,Blunden05}, do possesses this property when monopole form factors are used. However, as seen in Fig.~\ref{effects_from_vertex_paramerter_ff}, such a feature is not observed in our results for TPE-$\Delta$. They appear to either rise or decrease rapidly as  $\epsilon \rightarrow 1$, which
look surprising or even "pathological". It is not immediately clear to us why this is so. One possible explanation is that hadronic models such as ours, are not applicable when $s  \rightarrow \infty$ and  $\epsilon \rightarrow 1$. This is similar to the case that one does not expect the hadronic model to be reliable at large $Q^2$. At present, there exists no model calculation which is reliable at all scales. For example,
predictions of partonic calculations of \cite{Chen04,Afana05} are not expected to be reliable for small values of $\epsilon$. In
\cite{Kivel13},   the applicability of SCET  is stated to be restricted in the region of $\epsilon_{min} < \epsilon < \epsilon_{max}$, with $\epsilon_{min} \sim 0.42-0.60$ for $Q^2 \sim 3-6$ GeV$^2$.
A conservative estimate of the applicability of our hadronic model would be for
$W=\sqrt{s} \leq 3 - 4$ GeV. The corresponding range of   $\epsilon$ for $W \sim 3-4$ GeV and $Q^2 =1-4$ GeV$^2$ is depicted in Fig.~\ref{range of epsilon}. The vertical dashed line in Fig.~\ref{effects_from_vertex_paramerter_ff} correspond to a value of  $W = 3.5$ GeV, i.e., $\epsilon <0.904$ at $Q^2=3$ GeV$^2$. Hereafter we will restrict the comparisons of our predictions with  the experimental data at low $Q^2$ located in this region.

Contributions of TPE to $\delta$ 's have also been studied in the dispersion approach of \cite{BK12,BK14}.  For the case of  the contribution of TPE-N to $\delta_N$, our results, which are essentially the same as those obtained in \cite{Blunden05}, agree well with what are shown in Fig. 5 of \cite{BK12}. However, for $\delta_\Delta$, our results are considerably larger than the corresponding results obtained in \cite{BK12}. For example, the $\delta_\Delta$ 's at $Q^2=3$ GeV$^2$ shown by the thick dash-dotted line in Fig. 5 of \cite{BK12} is only about half of our results. In addition, we further find that $\delta_\Delta$ remains substantially smaller than $\delta_N$ at large momentum transfer $Q^2 \sim 6$ GeV$^2$ which is at variance with the findings of \cite{BK14}. The dispersion relation (DR) calculations  of \cite{BK12,BK14} for the TPE-$\Delta$ amplitude are based on the following three requirements. Namely, (i) it has no singularities except the branching point
at $s=(M_\Delta + m_e)^2$, (ii) \textcolor{black}{its branch cut discontinuity is $2iIm M^{(\Delta)} $ with
$M^{(\Delta)}=M^{(a,\Delta)}+M^{(b,\Delta)}$ as given by Eq. (\ref{eq:box-amp})}, and (iii) it vanishes as $s  \rightarrow \infty$. A close
look at the amplitude of $M^{(a,\Delta)}$ given in Eq. (\ref{eq:box-amp}) and the corresponding one for the crossed box diagram, clearly
indicates that the requirements of (i) and (ii) are satisfied except the $\Delta$ form factors employed are different from those  used in
\cite{BK12,BK14}, which are not expected to be responsible for the marked difference found in the above. The biggest difference between our
calculation and those of \cite{BK12,BK14} most likely lies in condition  (iii). This point remains to be further investigated.

\subsection{$\Delta(1232)$ contributions to the unpolarized cross section}

\begin{figure*}[htbp]
\center{
\includegraphics[width=1\textwidth]{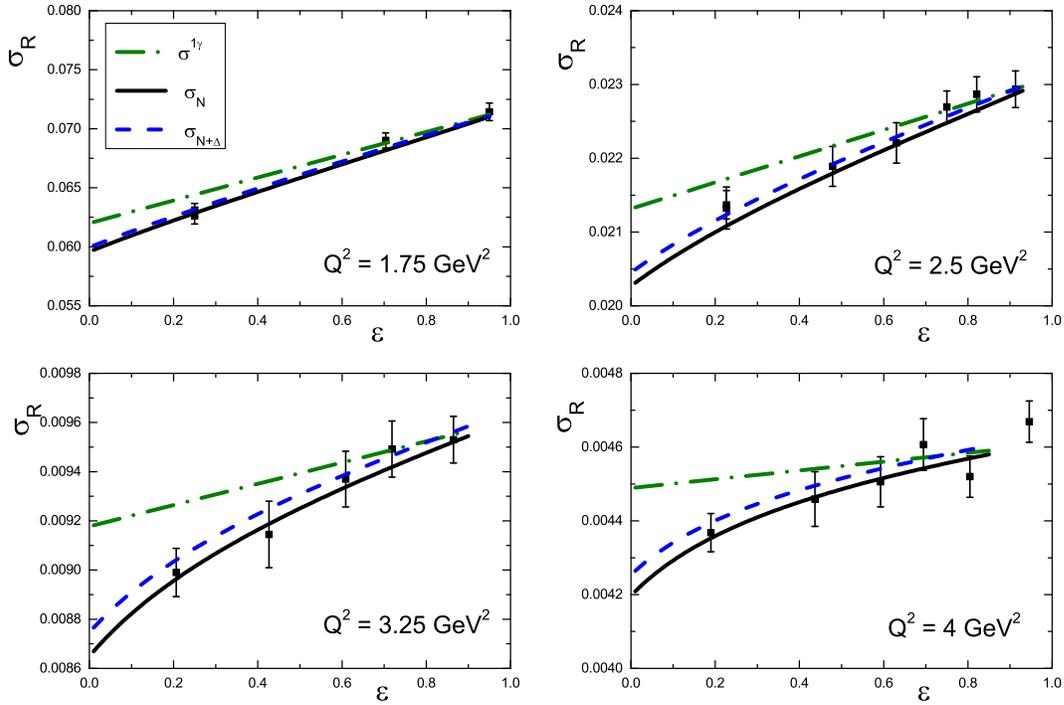}
\caption{The
unpolarized cross section as function of $\epsilon$ at fixed $Q^2$.
The (olive) dash-dot curves denote the Born cross sections, the (black)
solid,  and (blue) dashed curves refer to the cross sections
including only TPE-N and TPE-N plus TPE-$\Delta$. The theoretical
Born cross sections $\sigma^{1\gamma}$ are obtained as explained in
the text. Data are from \cite{Andivahis94}. } \label{CrossSection94}
}
\end{figure*}

\begin{figure*}[htbp]
\center{
\includegraphics[width=1\textwidth]{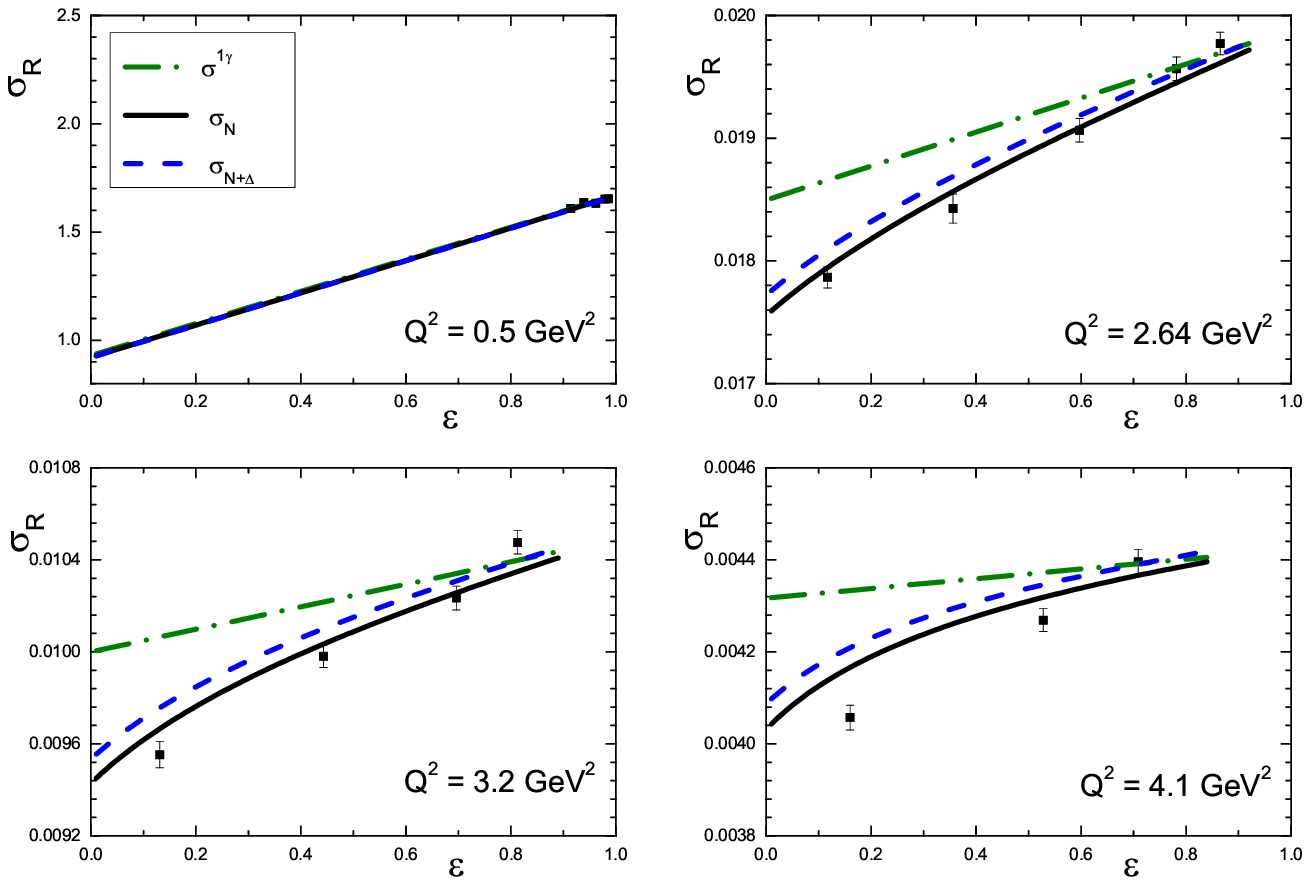}
\caption{Notations same as in Fig. \ref{CrossSection94}. Data are
from \cite{Qattan06}.} \label{CrossSection05}
}
\end{figure*}

In this subsection, we will compare our predictions with  only two
representative sets of data measured in 1994
\cite{Andivahis94,Arrington03} and 2006 \cite{Qattan06}, called as
data94 and data06, respectively. We do not consider the 1994 data of
\cite{Walker94} here as its feature is rather similar to that of
data06. The cross section arised from one-photon exchange,
$\sigma^{1\gamma}$, will be determined as follows. We first fix the
values of $R$ obtained from polarization experiments \cite{Jones00,Gayou02},
$R  = 1-0.13(Q^2-0.04)$. \textcolor{black}{As discussed in the last section, TPE
corrections to the cross sections are expected to be small and negligible, if not outright vanishing, when $\epsilon\rightarrow 1$.} Accordingly, we choose, with the simple least squares method, to fit the experimental reduced cross sections in the $\epsilon > 0.7$ region with the OPE expression of  Eq. (\ref{diffCr}) to determine $G_M$. It leads to  $G_M = (0.249, 0.146,$ $0.0958, 0.0670)$ at $Q^2 = (1.75, 2.5, 3.25, 4.0)$ GeV$^2$, and $G_M = (0.964, 0.136,$ $0.100, 0.0657)$ at $Q^2 = (0.5,$ $2.64, 3.2, 4.1)$ GeV$^2$, for data94 and data06, respectively. It should be pointed out that the theoretical reduce cross sections are sensitive to the values of $G_M$, especially at large $Q^2$ region. This is why we retain up to three significant digits in the above expressions. The resulting $\sigma^{1\gamma}$ $'$s, obtained from fitting to the data94 and data06 as explained above and represented by the (olive) dash-dotted curves are shown in Figs. \ref{CrossSection94} and \ref{CrossSection05} , respectively.

The cross sections including TPE contributions are evaluated as
$\sigma^{1\gamma}$ multiplied by the corresponding theoretical TPE
corrections via Eqs. (\ref{def-delta}, \ref{delta-def}). We mention
that our results including only TPE-N to be presented below are
consistent with those obtained in \cite{Blunden05}. \textcolor{black}{Note that a different choice of the values for $G_M$
will simply shift all three curves shown in each panel of Figs. (\ref{CrossSection94}) and (\ref{CrossSection05}) by a common factor of $G^2_M$, since $\bar\delta_N$ is fairly insensitive to the nucleon form factors as long as they are realistic, as found in Ref. \cite{Blunden05}.}

\subsubsection{1994 data set of Andivahis {\it et al.} }
The   unpolarized cross sections of data94 at $Q^2=(1.75, 2.5,$
$ 3.25,4.0)$ GeV$^2$ are denoted  in Fig. \ref{CrossSection94} by  (black) squares.
  The (black) solid curves, labeled as $\sigma_N$,
 correspond to the predictions  including
corrections of  TPE-N  only. It is seen that corrections from TPE-N
bring down the predictions of $\sigma^{1\gamma}$ to agree rather
well  to the data, especially for small $\epsilon$.

Further inclusion of TPE contributions arising from $\Delta$
intermediate states, labeled as $\sigma_{N+\Delta}$, are shown by
(blue) dashed curves. The difference between  (black) solid  and
(blue) dashed curves would then represent the contributions of
TPE-$\Delta$. The effect of  TPE-$\Delta$  clearly is smaller than
that of TPE-N and has opposite sign. It is seen that
$\sigma_{N+\Delta}$ does not improve the agreement between data and
$\sigma_N$ except for larger values of $\epsilon$ and $Q^2$.

\subsubsection{2006 data set of Qattan {\it et al.} }
The high precision super-Rosenbluth data set data06  are from
\cite{Qattan06}. The measured unpolarized cross sections at
$Q^2=0.5,$ $\,2.64, \,3.2, \,4.1$ GeV$^2$ are shown in Fig.
\ref{CrossSection05} and denoted by (black) squares.   Again the
(black) solid curves, labeled as $\sigma_N$,
 correspond to the predictions  including
corrections of  TPE-N  only and are seen to bring down the
predictions of $\sigma^{1\gamma}$ to agree rather well  with the
data, especially for small $\epsilon$. In contrast to the case with
data94, discrepancy between data and $\sigma_N$ begins to develop
with increasing $\epsilon$ and higher $Q^2$, and becomes substantial
for $\epsilon > 0.8$ and $Q^2 > 3.2$ GeV$^2$.

As with data94, TPE-$\Delta$ contributions are seen to be smaller in
magnitude   and have opposite sign with TPE-N such that
$\sigma_{N+\Delta}$,   denoted by (blue) dashed curves in
Fig. \ref{CrossSection05}, move $\sigma_N$ back toward $\sigma^{1\gamma}$ and
the nice agreement between data and $\sigma_N$ for  $\epsilon <0.7$
and  $Q^2 \leq 3.2$ GeV$^2$ is lost. However, for $Q^2 \geq 3.2$ GeV$^2$
and $\epsilon >0.8$, TPE-$\Delta$ actually is beneficial to bridge
the difference between data and $\sigma_N$.

The discussions presented in the above lead to the following
conclusion. Namely, contribution of TPE-$\Delta$ is smaller than
that of TPE-N and with opposite sign. For data94, TPE-$\Delta$
contribution, in most cases, brings our model predictions to agree
well with the data. For data06, TPE-$\Delta$ contribution is
beneficial only in region with larger values of $\epsilon$. However,
in the region with small values of $\epsilon$, TPE-$\Delta$
contribution move $\sigma_N$ away from the data.

\subsubsection{The uncertainties of TPE corrections from $\Delta(1232)$ }
\textcolor{black}{
There are two kinds of uncertainties in the above discussions within our model calculation. The first is from the uncertainties of the input parameters $g_1,g_2$ and $g_3$. From Fig. \ref{effects_from_vertex_paramerter_ff}(b), we see that the contribution from $g_3$ is small, so we will just focus on the uncertainties incurred from $g_1$ and $g_2$.  From Eq. (\ref{relation-gG}), it is seen that the uncertainties of $g_1$ and $g_2$ are almost equal and proportional to that of $G^*_M$, since  $R_{EM}$ is small in the low $Q^2$ region of our interest.  The  experimental uncertainty of $G^*_M$ is about $1\%$ . This uncertainty will give rise to an about $2\%$ global uncertainty of the TPE-$\Delta$ corrections $\delta_{\Delta}$, and leads to a correction less than $0.1\%$ to the extracted $R's$.}

\textcolor{black}{
The second uncertainty is associated with the form factors at finite $Q^2$ region. It can be estimated from   Figs. \ref{FFs-Delta} and  \ref{effects_from_vertex_paramerter_ff}(b). In Fig. \ref{FFs-Delta}, we see the two form factors (KBMT and ZY) are very different at finite $Q^2$, while their TPE corrections shown in Fig. \ref{effects_from_vertex_paramerter_ff}(b) are not much different when $g_3$ is set to zero. We can hence expect that a $300\%$ difference of $G_M^{*}$ at $Q^2=3$GeV$^2$ will result in   $100\%$ difference in the corresponding TPE-$\Delta$ corrections. Since the $G_M^{*}$ we use is very close to the experimental one, we expect this uncertainty to be about only at most a few percent. Since the TPE-$\Delta$ corrections are much smaller than TPE-N, this uncertainty will give an even smaller contribution to the uncertainty of the full TPE corrections.
}

\subsection{$\Delta(1232)$ contributions to the extracted $R$ in LT method}
We now turn to the correction of TPE to values of $R$ extracted from
LT (Rosenbluth) method. In the literature, there are two methods
proposed for such a determination. The first one \cite{Blunden03}
parameterizes $  1+\Delta_{un}=a(1+b\epsilon)$ and the corrected $R$
is taken as $\sqrt{R_0^2-b/B}$ where $R_0$ is the extracted $R$
without the inclusion of TPE corrections and $B=1/{\mu_p^2\tau}$.
The second method \cite{Arrington07} applies the TPE corrections to
the experimental data  and then fit  the corrected data sets with
Eq. (\ref{diffCr}). Namely, we divide the experimental cross
sections by the factor of $(1+\Delta_{un})$ as in Eq.
(\ref{def-delta}) and determine the slope via Eq. (\ref{diffCr}). We
call these two methods as linear parametrization and direct fitting
method, respectively. We have applied both methods on the data
measured in 1994 \cite{Walker94, Andivahis94}, which have large
error bars, and the data of the recent high-precision super-Rosenbluth experiment \cite{Qattan05,Qattan06} measured in
2005 at Jlab. Both methods yield quantitatively similar results.
Accordingly, we'll present only results obtained with the fitting
method with data obtained from each single  $Q^2$ analysis.

\begin{figure*}[htbp]
\center{
\includegraphics[width=1\textwidth]{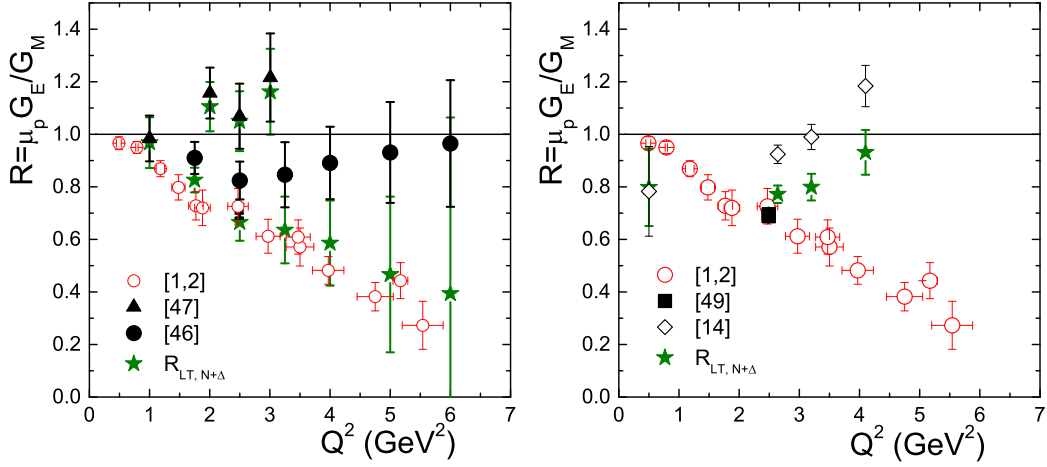}\caption{The
extracted $R's$ from  the LT method after the removal of TPE effects
and via direct fitting, are shown by (green) stars and labeled as
$R_{LT, N+\Delta}$. Solid triangles, solid circles, and open rhombi
denote the values of $R_0$ extracted by the experimentalists, i.e.,
without considering the TPE effects. The PT data are denoted by open
circles \cite{Jones00,Gayou02} and solid squares \cite{Meziane11}. Left
panel: results obtained with the 1994 data of \cite{Walker94,
Andivahis94}. Right panel: results obtained with the 2006 data of
\cite{Qattan06}. } \label{RLT-extracted}
}
\end{figure*}

Our results for the TPE corrections to the values of $R$ extracted
from LT method, with the  data of  \cite{Walker94, Andivahis94} and
\cite{Qattan06}, are presented in  Fig. \ref{RLT-extracted}, and
compared with $R's$ extracted from PT measurements
\cite{Jones00,Gayou02,Meziane11} as denoted by open circles and solid
squares. The solid triangles, circles, and open rhombi, correspond
to the values of $R_0$ extracted by the experimentalists which did
not include any TPE effects. The (green) stars represent our
extracted values of $R$ by fitting method after removing the effects
of TPE, including both TPE-N and TPE-$\Delta$, as prescribed by our
model where the  error bars for $R_{LT,N+\Delta}$ are estimated with only
the statistical and point-to-point uncertainty presented in \cite{Arrington03, Qattan06}  considered. \textcolor{black}{The above comparison for $Q^2 > 4$ GeV$^2$ should be taken with caution since our hadronic model calculation might not be reliable in such high $Q^2$ region. }

From the left panel of Fig. \ref{RLT-extracted}, we see that the TPE
effects prescribed by our model can almost explain the discrepancy
in the values of $R$ as extracted from LT and PT methods, as far as
only the LT data of \cite{Andivahis94} are considered. However,
substantial discrepancy remains in the case of the LT data of
\cite{Walker94,Qattan06} even though the TPE effects do help to
explain part of the discrepancy.

From the discussions in the last subsection and here, more cross
section experiments will be very helpful to shed light on how to
further improve model calculation.

\subsection{$\Delta(1232)$ contribution to the ratio $R^{\pm}$ between the positron-proton and electron-proton cross sections }
The amplitudes for the positron-proton (${\it e^+p}$) and
electron-proton (${\it e^-p}$) scatterings can be written as
$T^{(\pm)}=\pm T_{1\gamma}+T_{2\gamma}$, where  $(\pm)$ correspond
to the charge of positron and electron, and $T_{1\gamma}$ and
$T_{2\gamma}$ denote the scattering amplitudes with $1\gamma$ and
$2\gamma$ exchanged, respectively.  We then have ratio between the
unpolarized cross sections of (${\it e^+p}$) and (${\it e^-p}$)
elastic scattering given as,
\begin{eqnarray}
R^{(\pm)}\equiv \frac {\sigma(e^+p)}{\sigma(e^-p)} \simeq 1+
4Re\left(\frac{T_{2\gamma}}{T_{1\gamma}}\right)=1-2\Delta_{un},\label{e+/e-}
\end{eqnarray}
where $\sigma(e^\pm p)$ refer to the unpolarized cross sections of
$e^\pm p$ elastic scatterings. Thus measurements of the ratio of
$e^+p$ and $e^-p$ cross sections provide a direct probe of the real
part of the TPE amplitude.

Earlier measurements on $R^\pm$, limited by the low intensity of $e^+$  beams and hence
with large error bars, have been compiled in \cite{Arrington04,Arrington11}. Three experiments
have recently been undertaken. Two of them have finished data taking \cite{VEPP3-2012,VEPP3-2014,Moteabbed13} with
preliminary data published while the third is expected to run soon \cite{Kohl14}.
In the followings, we will compare our predictions with the published data of \cite{VEPP3-2012,VEPP3-2014,Moteabbed13}.

\begin{figure*}[htbp]
\center{
\includegraphics[width=1\textwidth]{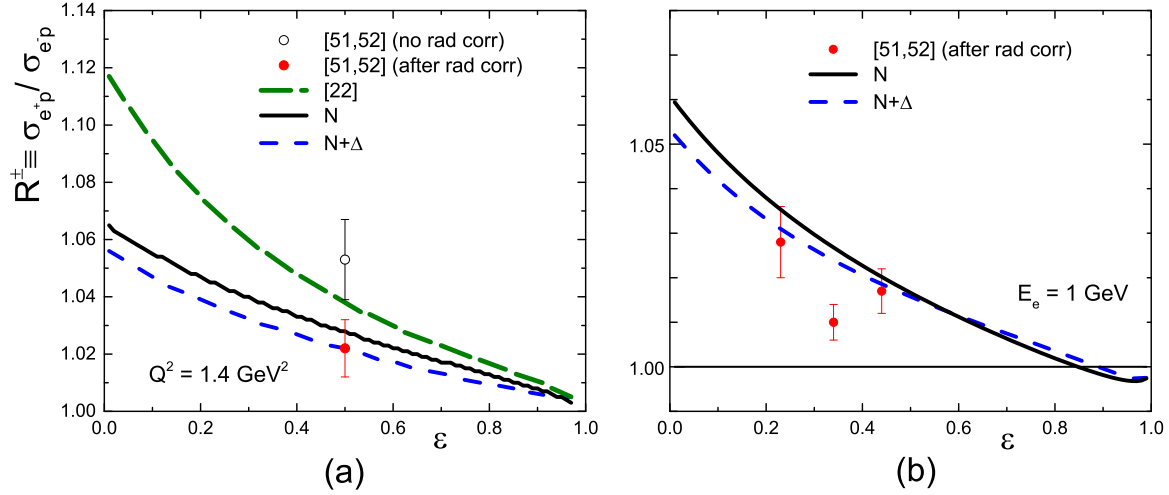} \caption{TPE
corrections to the ratio
   $R^{\pm}$. The (black)  solid and (blue)
dashed curves, refer to predictions with  TPE corrections from the $N$ only, and
$N+\Delta$ intermediate states included, respectively.   The preliminary
experimental data are taken from \cite{VEPP3-2012,VEPP3-2014}, with open and solid
circles corresponding to data before and after the radiative
corrections are applied. (a) $R^{\pm} $ vs. $\epsilon$ at $Q^2=1.4$
GeV$^2$. The dash-dotted curve, is the prediction of a
model-independent parametrization II of TPE corrections of
\cite{Chen07}. (b) $R^{\pm} $ vs. $\epsilon$ at incident electron
lab energy $E_e = 1$ GeV.} \label{Rpm}
}
\end{figure*}

Our predictions for $R^{(\pm)}$, labelled as $N$ and $N+\Delta$ and
denoted by (black) solid and (blue) dashed lines, corresponding to
results with
the contributions of   TPE-N   and TPE-N plus TPE-$\Delta$
  are shown in Fig. \ref{Rpm}, respectively,
and compared to the
preliminary experimental data of VEPP-3 \cite{VEPP3-2012,VEPP3-2014}. The
open and solid circles denote the data before and after the
radiative corrections are applied. In Fig. \ref{Rpm}(a) $R^{(\pm)}$
vs. $\epsilon$ at $Q^2=1.4 $ GeV$^2$ is depicted, where the
prediction of fit II of a model-independent parametrization of TPE
effects in \cite{Chen07}, are also shown.  We have chosen to present
 the data and our predictions for $R^{(\pm)}$ vs. $\epsilon$ at
fixed $Q^2=1.4$, instead of $R^{(\pm)}$ vs. $\epsilon$ at fixed
incident electron lab energy $E_e=1.6$ GeV as was done in the left
panel of Fig. 1 in \cite{VEPP3-2012,VEPP3-2014} is because a CLAS experiment at the
same $Q^2$ has recently finished data taking and being analyzed
\cite{Weinstein13}. Fig. \ref{Rpm}(b) shows $R^{\pm} $ vs.
$\epsilon$ at incident electron lab energy $E_e = 1$ GeV.

It is seen in Fig. \ref{Rpm} that, in general, our results for
$N+\Delta$ agree with the preliminary data of VEPP-3 well except for
the point at $E_e = 1$ GeV and $\epsilon=0.34 \,(Q^2=0.90$ GeV$^2)$.
The inclusion of $\Delta$ in the intermediates states in the TPE
diagrams is also seen to somewhat improve  the agreement with the
data. The effect of TPE associated with $\Delta$ excitation on
$R^{\pm}$, though small at large $\epsilon$, becomes substantial at small
$\epsilon$. We also find that it is very important to use the
correct $\gamma N\Delta$ vertex function as employed in this
investigation in this kinematical region.

The good agreement between our  prediction and the data for $R^\pm$
is encouraging and indicates that the real part of $T_{2\gamma}$
prescribed by our model of TPE might be a reasonable one, at least
in the small $Q^2$ region.

\begin{figure}[htbp]
\center{
\includegraphics[width=0.45\textwidth]{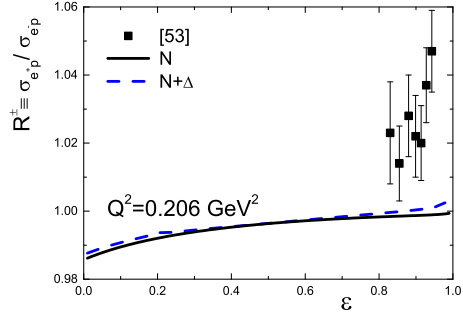}
\caption{TPE corrections  to
$R^{\pm}$ at $Q^2 = 0.206$ GeV$^2$. Notations same as  Fig. \ref{Rpm} and  the
data are  from  \cite{Moteabbed13}.} \label{Rpm-QQ0206}
}
\end{figure}

We next compare our predictions with the recent CLAS data listed in Table II of \cite{Moteabbed13} at $Q^2 = 0.206 $
GeV$^2$ as shown in Fig. \ref{Rpm-QQ0206}, with the same notation as in Fig.  \ref{Rpm}. The large luminosity-related
systematic uncertainty of 0.05 given there are not included in the figure.

We see considerable discrepancy between our prediction and the data if the large luminosity-related systematic uncertainty is not included.  It will be interesting to see whether such discrepancy persists after the large luminosity-related systematic uncertainty is reduced from the experiment.  Here we see our prediction with TPE-N
approaches one when $\epsilon \rightarrow 1$ as expected from the argument presented at the end of
Sec. III-A. The results with TPE-$\Delta$ included, however, do begin to increase near $\epsilon = 1$
as hinted by the data. This brings up an interesting question. Namely, whether our results for TPE-$\Delta$
is a realistic one or the uncertainty of the beam luminosity in the experiment of \cite{Moteabbed13}
will eventually bring the data down to one near $\epsilon = 1$.

\subsection{$\Delta$ contribution to the single spin asymmetries $B_n$ and $A_n$}
We now turn to the effect of TPE in the single spin asymmetries
$B_n$ and $A_n$. Since both  vanish within OPE approximation because
of the time reversal invariance, they provide  direct  access to the
TPE amplitude. However, in contrast to $R^{\pm} $ discussed in the
last subsection which probes the real part of $T_{2\gamma}$,  $B_n$
and $A_n$ are related to the imaginary part of the of the TPE
amplitude instead.

\subsubsection{Beam-normal single spin asymmetries $B_n$}

For a beam polarized  perpendicular to the scattering plane, the single spin asymmetry is defined
as
\begin{eqnarray}
B_{n} \equiv \frac{\sigma^{\uparrow}_e - \sigma^{\downarrow}_e}{\sigma^{\uparrow}_e + \sigma^{\downarrow}_e},
\label{Bn-defination}
\end{eqnarray}
where $\sigma^{\uparrow}_e(\sigma^{\downarrow}_e)$ denotes the cross
section for   unpolarized proton target and   electron beam spin
parallel (antiparallel) to the vector $\hat n$ normal to the
scattering plane,
\begin{eqnarray}
\hat n=\frac{\vec p_1 \times \vec p_3}{|\vec p_1 \times \vec p_3|}.
\label{def-normal}
\end{eqnarray}
%Generally, the $B_n$ is proportional to the imagine part of TPE amplitude and the mass of electron\cite{Chen07}.

%Respectively, And Is Given By,
%\Begin{Eqnarray}
%B_{N}=\Frac {2Im(T_{1\Gamma}T^*_{2\Gamma})}{|T_{1\Gamma}|^2}.
%\Label{Bn-Defination}
%\End{Eqnarray}

It is a challenging task to measure $B_n$ because to polarize an
ultrarelativistic electron in the direction normal to its momentum
involves a suppression factor of $m_e/E_e$ which is of the order of
$10^{-4}-10^{-3}$ for $E_e$ of the order of GeV.  This
type of difficult experiments
\cite{Wells01,Maas05,Capozza07,Armstrong07,Armstrong11} have been carried out as
by-product of the intensive effort to measure the nucleon strange
form factors from the parity-violating asymmetry of the elastic
electron-proton scattering \cite{Androic10}. The TPE and $\gamma
Z$-exchange corrections to the parity-violating asymmetry have been
studied in \cite{Afanasev05,Zhou07,Nagata09}.

As elaborated in \cite{Pasquini04}, the imaginary part of the TPE
amplitude can be related, via unitarity, to the doubly virtual Compton scattering tensor on the nucleon with all possible intermediate hadronic states to be {\it on-shell}. In \cite{Pasquini04}, they considered only the
contributions of $\pi N$ intermediate states by modeling the doubly
virtual Compton scattering tensor in terms of the $\gamma^* N
\rightarrow \pi N$ amplitude. In our calculations of $B_n$ and
$A_n$, we will assume that in the resonance region, $\pi N$
intermediate states are saturated by the excitation of $\Delta$ with
a realistic decay width.  We follow the recipe of \cite{Oset08} to
account for the effect of the $\Delta$ width on  $B_n$ (and
similarly $A_n$ in the following subsection) as follows, with the
familiar Breit-Wigner form of constant width $\Gamma_\Delta = 116$
MeV,

\begin{eqnarray}
B_n &=& \int _{M_\Delta-2\Gamma_\Delta}^{M_\Delta
+2\Gamma_\Delta}B_n( M_D)\rho(M_D)dM_D, \nonumber \\
\rho(M_D) &=&
-\frac{1}{\pi}Im\left[\frac{2M_D}{M_D^2-
M_\Delta^2+iM_\Delta\Gamma_\Delta}\right], \label{Bn-width}
\end{eqnarray}
where $B_n(M_D)$ is given by Eq. (\ref{Bn-defination}) with the mass
of $\Delta$, $M_\Delta$, replaced by $M_D$.

\begin{figure*}[htbp]
\center{
\includegraphics[width=1\textwidth]{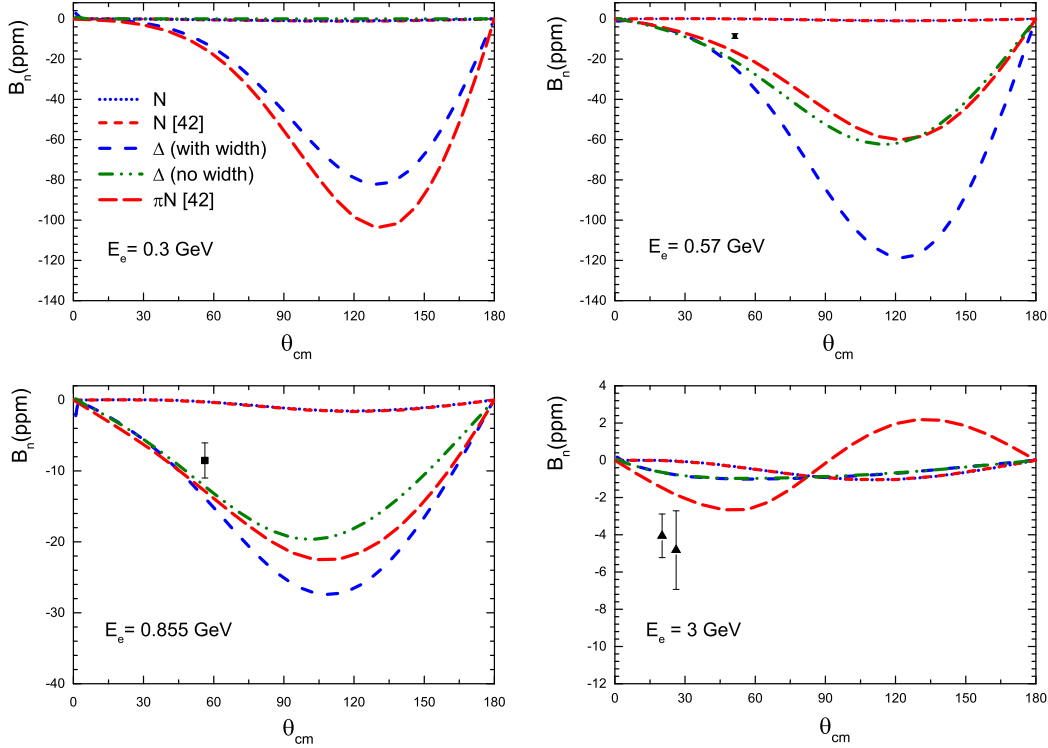}
\caption{Our
predictions for the effects of TPE-N and TPE-$\Delta$, with and
without $\Delta$ width, on $B_{n}$ vs. $\theta_{CM}$ at  $E_e=0.3,
0.57, 0.855, 3 \,$GeV as compared with results obtained in
\cite{Pasquini04}. The notation for the various curves are explained
in the left top panel. The (red) dot-dashed curves denote the
results of \cite{Pasquini04} with $\pi N$ continuum in the
intermediate states. The data are from \cite{Maas05,Armstrong07,Armstrong11}.}
\label{Bn1}
}
\end{figure*}
Our predictions for $B_n$ vs. CM angle $\theta_{cm}$ at four
electron energies $E_e = 0.3, 0.57, 0.855, 3 \,$ GeV are presented
in Fig. \ref{Bn1} and compared with results obtained in
\cite{Pasquini04}, where $\gamma^* N\rightarrow\pi N$ amplitude is
taken from a phenomenological analysis of electroproduction
observables \cite{Tiator01}. Both calculations  obtain very small
contributions from TPE with only nucleon in the intermediate states
as indicated by (red) short-dashed and (green) dotted lines,
respectively. Our results for contributions from $\Delta$ without
and with width are given by (blue) dashed and (green) dot-dot-dashed
lines, while the contributions from $\pi N$ intermediate states as
estimated by \cite{Pasquini04} are denoted by (red) dot-dashed
lines.

At $E_e = 0.3 \,$GeV in the upper left panel of Fig. \ref{Bn1}, it
is seen that the contribution from $\Delta(1232)$ intermediate
states is zero if $\Delta$ is treated as a stable particle, i.e.,
with the $\Delta$ width taken to be zero. This can be understood as
follows. Namely, $B_n$ is related to the imagine parts of the TPE
amplitude which would receive contributions only from {\it on-shell}
intermediate states. For the $e \Delta$ intermediate states,
on-shell condition leads to a threshold energy for the electron $E_e^{thr}$,
\textcolor{black}{
\begin{eqnarray}
E_e^{thr} = \frac{M_\Delta^2-M_N^2+2M_{\Delta}m_e}{2M_N}\approx 0.34 GeV, \label{condtion-Delta}
\end{eqnarray}}
In the calculation of \cite{Pasquini04}, the inelastic intermediate
states are taken as $\pi N$ and the on-shell conditions result in a
threshold value of $E_e^{thr} = 0.151$ GeV which is smaller than
$0.3$ GeV. This is why \cite{Pasquini04} would obtain nonvanishing
result for $B_n$ in the case of $E_e=0.3$ GeV, as shown in the upper
left panel of Fig. \ref{Bn1}. It is seen that the effect of the
$\Delta$ width is substantial but begin to decrease as energy
increases to pass over the region dominated by the $\Delta$. Note
that the vertical scales in the lower two figures are different from
the upper two.

For $E_e = 0.3, 0.57, 0.855 \,$ GeV, our results show similar
angular dependence as those obtained in \cite{Pasquini04} but the
absolute magnitude of our result at $E_e = 0.57 \,$ GeV is
considerably larger. The two data points at $E_e = 0.57, 0.855
\,$GeV  come from \cite{Maas05} and their absolute magnitudes are
smaller than the predictions of ours and those of \cite{Pasquini04}.
At $E_e =3 \,$GeV, the absolute magnitudes of our results are much
smaller than experimental data \cite{Armstrong07,Armstrong11} and also show very
different behavior with the results in \cite{Pasquini04}. This  can
be understood naturally as the center of mass energy $\sqrt{s}$
reaches about $4\,$GeV, where the higher resonances, not considered
in our model, will dominate.

\begin{figure*}[htbp]
\center{
\includegraphics[width=1\textwidth]{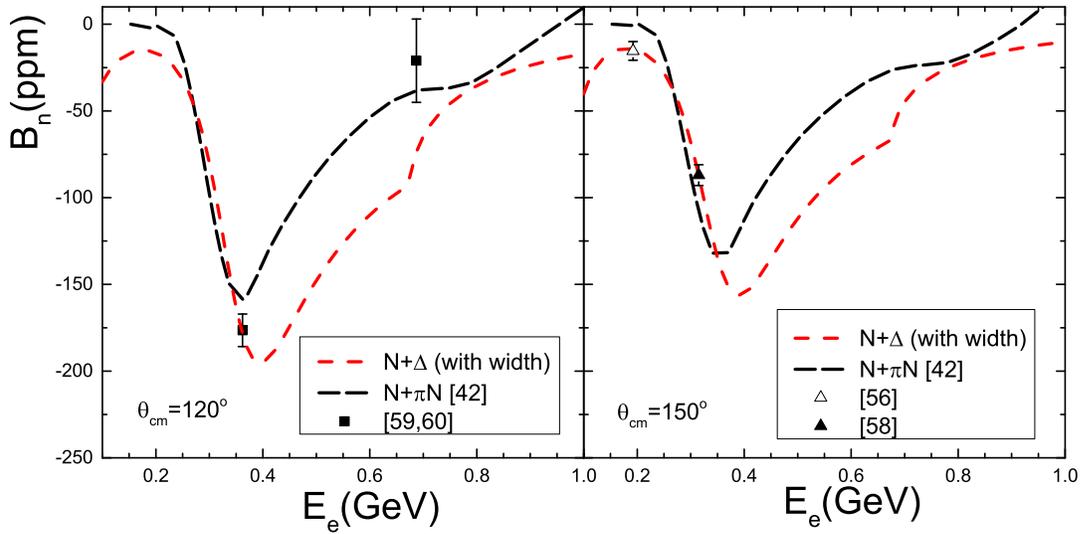} \caption{Our
predictions for $B_n$ vs. $E_e$ at $\theta_{cm} = 120^\circ,
150^\circ$, with contributions coming from nucleon and $\Delta$
together and denoted by (red) dashed lines,  are  compared with
results obtained in \cite{Pasquini04}, as given by (black)
dot-dashed lines. The data are from
\cite{Wells01,Capozza07,Armstrong07,Armstrong11}.} \label{BnEe}
}
\end{figure*}

In Fig. \ref{BnEe}, our predictions for the variations of $B_n$ {\it
w.r.t.} electron energy $E_e$ at $\theta_{cm} = 120^\circ$ and
$150^\circ$ are shown, and compared with the corresponding results
of \cite{Pasquini04}, denoted by (red) dashed  and (black)
dash-dotted lines, respectively, and the experimental data
\cite{Wells01,Capozza07,Armstrong07,Armstrong11}. The kinks seen in our
predictions around $\theta_{cm}$ arise from the competition between
the contribution of the mass $M_\Delta$ and the width
$\Gamma_\Delta$  as explained earlier. It is interesting to see that
our predictions agree with the data better than those of
\cite{Pasquini04} except for one data point at $\theta=120^\circ$
with $E_e \sim 0.7$ GeV.

\subsubsection{Target-normal single spin asymmetries $A_n$}

The target-normal spin asymmetry $A_n$ is defined as
\begin{eqnarray}
A_{n} \equiv \frac{\sigma^{\uparrow}_{p}-\sigma^{\downarrow}_{p}}{ \sigma^{\uparrow}_{p}+ \sigma^{\downarrow}_{p}},
\label{An-defination}
\end{eqnarray}
where $ \sigma^{\uparrow\downarrow}_{p}$ are the corresponding cross
sections of $e(p_1)$ $p^{\uparrow\downarrow}(p_2)\rightarrow
e(p_3)p(p_4)$  with the polarization vector of the target proton
normal to the scattering plane. To including the effects from the
width of the  intermediate  $\Delta$ for $A_n$, we use  similar
expression for   $B_n$ as given in Eq. (\ref{Bn-width})
\begin{eqnarray}
A_n &=& \int
_{M_\Delta-2\Gamma_\Delta}^{M_\Delta+2\Gamma_\Delta}A_n(
M_D)\rho(M_D)d M_D. \label{An-width}
\end{eqnarray}

\begin{figure*}[htbp]
\center{
\includegraphics[width=1\textwidth]{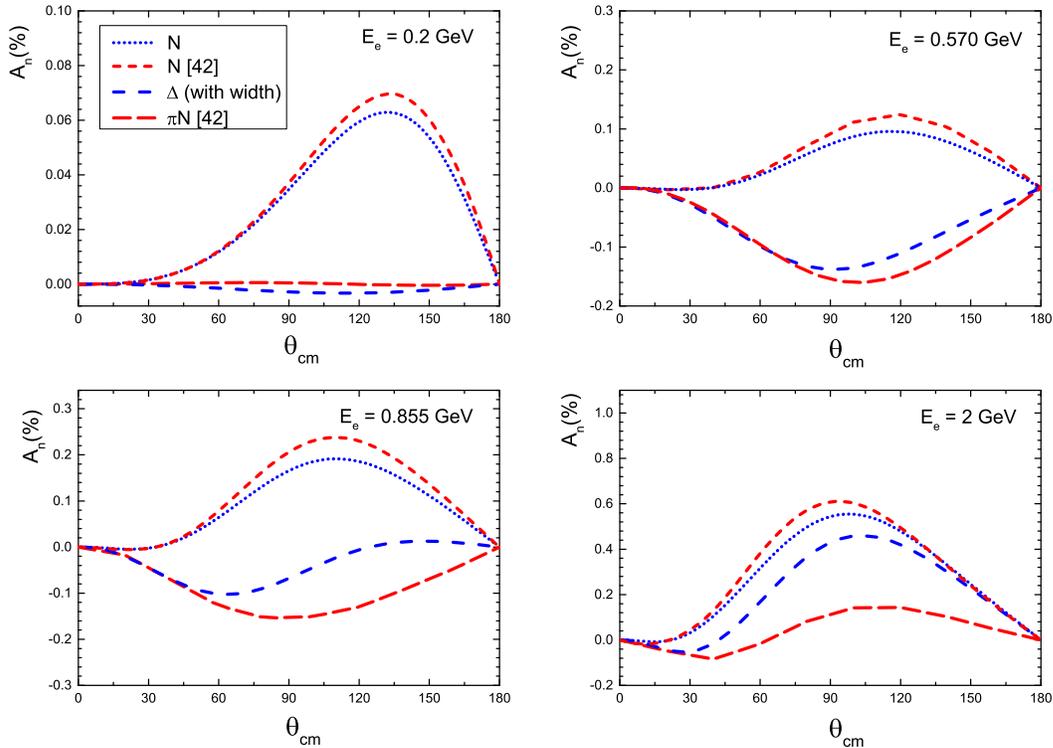} \caption{Our
predictions for $A_{n}$ vs $\theta_{cm}$ at fixed initial electron
 $E_e = 0.2, 0.570, 0.855, 2 \,$GeV. Notations same as in Fig. \ref{Bn1}.
 Results from \cite{Pasquini04}
are also shown for comparison.} \label{An}
}
\end{figure*}

Fig. \ref{An} shows our predictions for $A_n$ vs. $\theta_{cm}$, and
compared with the results of  \cite{Pasquini04}. It is seen that the
results coming from the nucleon intermediate states are similar
angular variation, though differ in magnitudes by $\sim 15 \%$. For
the inelastic contributions, at $E_e = 0.2, 0.57 \,$GeV cases, our
results are also very close to those obtained in \cite{Pasquini04}.
However, for  $E_e = 0.855$ and $2$ GeV, our results and those
obtained in \cite{Pasquini04} agree only at the small $\theta _{cm}$
and begin to differ at larger angle, say, for $\theta_{cm} >
30^\circ$ at $E_e = 2$ GEV, as in the case of $B_n$. The difference
lies not only on magnitude but also in angular dependence. It could
be attributed to the treatment of the $\Delta$ width and the
contributions from higher nucleon resonances.

\subsection{$\Delta$ contribution to the polarized variables $P_t, P_l,$ and $ R_{PT}$}

In the last five subsections, we are concerned only with the TPE
corrections to the unpolarized observables and single spin
asymmetries $A_n$ and $ B_n$. However, since the interest in TPE
effects arises from the discrepancy between the values of $R$
extracted from Rosenbluth separation (LT) and polarization transfer
(PT) methods, it is hence important that we also study the  TPE
corrections to the polarization observables $P_t, P_l$.

The TPE corrections to  $P_t, P_l$ was studied in a hadronic model
in \cite{Blunden05}. However, they only considered the correction of
TPE arising from $N$ intermediate states.  In the followings, we
present our predictions for the TPE corrections from both $N$ and
$\Delta$ intermediate states to   $P_t, P_l$ and compare them with
the data of a recent precise measurement  carried out at Jefferson
Lab in Hall C,  in the $\vec e + p\rightarrow e + \vec p$ elastic
scattering \cite{Meziane11}.

The longitudinal  and transverse  polarizations of the recoil proton
with a longitudinally polarized electron of helicity $\lambda$ are
given by
\begin{eqnarray}
\lambda P_{t,l}  \equiv
\frac{\sigma^{+}_{t,l}(\lambda)-\sigma^{-}_{t,l}(\lambda)}{\sigma^{+}_{t,l}(\lambda)+\sigma^{-}_{t,l}(\lambda)},
\label{Ptl}
\end{eqnarray}
where $\sigma^{\pm}_{t,l}(\lambda)$ denote the cross sections of
$e(p_1,s_1)p(p_2)\rightarrow e(p_3)p(p_4,  s_{t,l})$ with $s_{t,l}$
the corresponding transverse and longitudinal polarization vectors
(in the scattering plan) of the final proton
\cite{Tomasi-Gustafsson07,Ladinsky1992}. Namely, if we denote the
spin direction of the recoil proton in its rest frame as $\vec
\zeta$, then $\vec \zeta_l \parallel \vec p_4$ and $\vec \zeta_t
\parallel \hat x$, where $\hat x = \hat y \times \hat z$, with
unit vectors $\hat y$ in the direction of $\vec p_1 \times \vec p_3$
and $\hat z \parallel \vec p_4$. The superscripts + and - correspond
to the cases where $\vec \zeta_{l,t}$ are parallel or antiparallel
to $\vec p_4$ and $\hat x$, respectively. Note that $P_{t,l}$ is
independent of $\lambda$. We can also write
\begin{eqnarray}
\sigma^{\pm}_{t,l}(\lambda)=\frac{1}{2}\sigma_{un} (1 \pm \lambda
P_{t,l}), \label{Ptl-a}
\end{eqnarray}
where the unpolarized cross section is given by
\begin{eqnarray}
 \sigma_{un}&=&\frac{1}{2}(\sigma_{t,l}^{+}(+1)+\sigma_{t,l}^{+}(-1)+ \sigma_{t,l}^{-}(+1)+ \sigma_{t,l}^{-}(-1))\nonumber\\
&=& \sigma_{t,l}^{+}(+)+ \sigma_{t,l}^{-}(+)\nonumber\\
&=& \sigma_{t,l}^{+}(-)+ \sigma_{t,l}^{-}(-).\label{sigma-pm}
\end{eqnarray} \noindent The second and the third lines in the above equation
hold because parity conservation leads to $\sigma_{t,l}^{m}(\lambda)
= \sigma_{t,l}^{-m}(-\lambda)$.

In OPE approximation,
\begin{eqnarray}
P_t^{1\gamma} &=&
-\frac{1}{\sigma_R}\sqrt{\frac{2\epsilon(1-\epsilon)}{\tau}}G_EG_M, \nonumber \\
P_l^{1\gamma} &=&  \frac{1}{\sigma_R}\sqrt{(1-\epsilon^2)}
G_M^2,
\end{eqnarray}
which leads to the well-known result of Eq. (\ref{polR}). \noindent
The TPE and other higher-order corrections to $P_t,P_l$ and $R_{PT}$
are defined as, in analogous to Eq. (\ref{def-delta}),
\begin{eqnarray}
P_{t,l} = P_{t,l}^{1\gamma}(1+\delta P_{t,l} ), \hspace{1.0cm}
R_{PT} = R_{PT}^{1\gamma} (1+\delta R_{PT}) , \label{delta-RPT}
\end{eqnarray}
where $R_{PT}^{1\gamma}\equiv \mu_p G_E/G_M$ would be value of
$R_{PT}$ if all higher-order corrections beyond OPE, including TPE,
are negligible.

Since we consider here only the higher-order effects up to TPE, we
will equate $P_{t,l}= P_{t,l}^{1\gamma+2\gamma}$ and $R_{PT}=
R_{PT}^{1\gamma+2\gamma}$, where the superscripts
 $1\gamma+2\gamma$ refer to  $P_{l,t}$'s evaluated
within   $1\gamma+2\gamma$ approximation. It is straightforward,
albeit tedious, to calculate $P_{t,l}^{1\gamma+2\gamma}$ according
to either Eq. (\ref{Ptl}) or Eq. (\ref{Ptl-a}). We mention that in the actual calculation, the IR divergences in the   $\sigma^{\pm}_{t,l}(\lambda)$'s have been subtracted as done in Ref. \cite{Blunden05}.

\begin{figure*}[hbtp]
\center{
\includegraphics[width=1\textwidth]{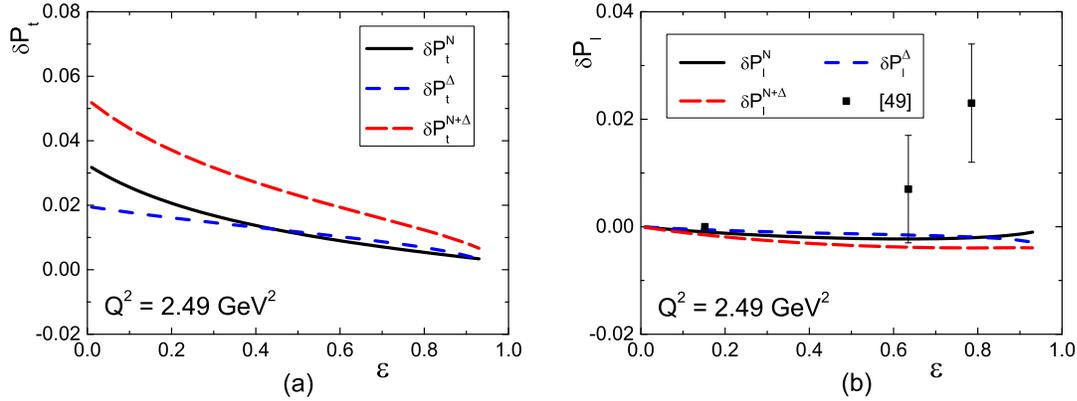} \caption{Our
predictions of the TPE corrections to $P_{t,l}$.  $\delta P^N_{t,l}$
and $\delta P^{\Delta}_{t,l}$, denoted by (black) solid and (blue)
dashed lines,
 refer to the corrections arising from $N$ and $\Delta$ in the intermediate states,
respectively.  The sum is represented by dash-dotted curves and data are
 from \cite{Meziane11} with $\delta P_l$
normalized at $\epsilon=0.152$.} \label{deltaPTR}
}
\end{figure*}

Our results at $Q^2=2.49 \,$ GeV$^2$ for the TPE corrections to
$\delta P_{t,l}$ are presented in Fig. \ref{deltaPTR},
 where contributions coming from $N$ and $\Delta$ in the intermediate states, are
 denoted by (black) solid and (blue) dashed lines, respectively, with their sum given by
 dash-dotted curves. The data for $\delta P_l$,  normalized at
$\epsilon=0.152$ by the experimentalists are from \cite{Meziane11}.
It is seen that the our predictions for TPE corrections remain small
for $\delta P_l$ throughout the entire region of $\epsilon$ and fall
considerably below the experiment  for  the two data points at
$\epsilon = 0.635, 0.785$ as shown in Fig. \ref{deltaPTR}(b).
For $\delta P_t$, the TPE corrections coming from $N$ and $\Delta$ are both
small but not negligible at small values of $\epsilon$ as seen in
Fig. \ref{deltaPTR}(a), with nucleon contribution larger than that
of the $\Delta$. However, both drop quickly for $\epsilon \ge 0.4$.

\begin{figure*}[hbtp]
\center{
\includegraphics[width=1\textwidth]{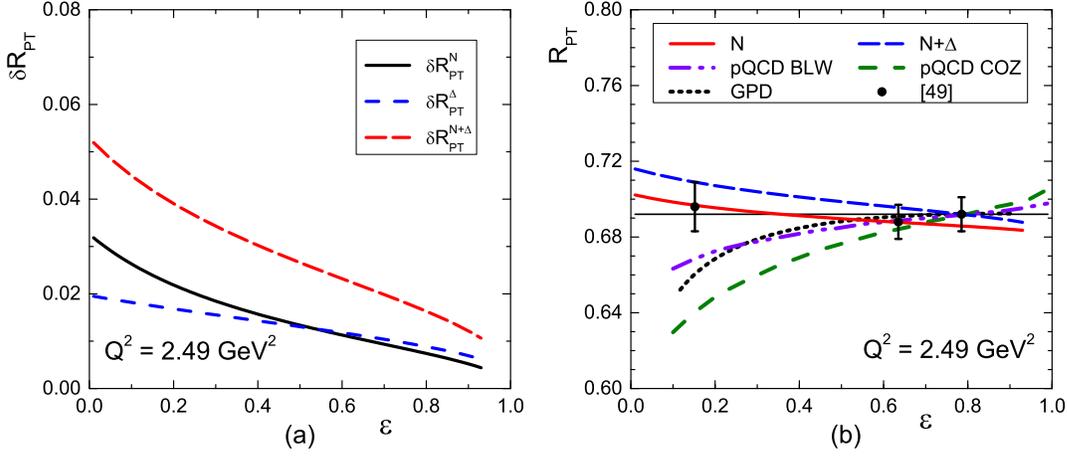}\caption{(a) Our
predictions for  $\delta R_{PT}$ with same notation as in Fig.
\ref{deltaPTR}.  (b) Our predictions for $R_{PT}$, together with
predictions of other theoretical calculations of \cite{Afana05}
(partonic) and \cite{Kivel09} (pQCD). Data are from
\cite{Meziane11}.} \label{deltaRPL}
}
\end{figure*}
Our results for $\delta R_{PT}$   are shown in Fig. \ref{deltaRPL}
(a) with the same notation as that of     Fig. \ref{deltaPTR}. It is
easy to see from Eq. (\ref{delta-RPT}) that   $\delta R_{PT}\simeq
\delta P_t - \delta P_l \simeq \delta P_t$ since $\delta P_l$ is
small. That's the reason
 $\delta R_{PT}$ behaves very similar to $\delta P_t$ of Fig.
\ref{deltaPTR}(a). In Fig. \ref{deltaRPL} (b), our results for
$R_{PT}$ are presented and compared with data of \cite{Meziane11},
as well as results of other theoretical calculations, including the
partonic \cite{Afana05} and pQCD \cite{Kivel09} ones. Please note
that we have normalized the $R_{PT}^{1\gamma}(1+\delta R_{PT})$ to be  equal to $0.692$ at $\epsilon= 0.785$ for all
model calculations except the (red) solid curve which is normalized with respect to the (blue) dash-dotted curve. This is different from what was done in \cite{Meziane11}.  It is seen that the prediction of the
TPE hadronic model calculation including only the nucleon
intermediate states does roughly reproduce the data but adding the
effect of the $\Delta$ shifts the curve upward by about $2\%$,
whereas all other calculations fail badly, especially at small
$\epsilon$ region.

More precision measurements of the polarization transfer observables similar to that of \cite{Meziane11} will be most helpful to understand, quantify, and characterize the two-photon-exchange mechanism in electron-proton scattering.

%The measurements of the polarization transfer observables by the
%GEp2$\gamma$ collaboration clearly pose a severe challenge to the
%theory of TPE effects.

\section{Conclusions}
We have revisited the question of the contributions of the
two-photon exchange associated with the $\Delta$ excitation, to
various observables, unpolarized as well as polarized, in the
elastic electron-proton scattering, in a hadronic model.   Three
improvements over previous studies are made in our calculations in
order to obtain a better estimate on this important mechanism in the
hope of gaining better insight on how to resolve the puzzling
discrepancy between the value of $R=\mu_pG_E/G_M$ extracted from LT
and PT measurements.

The three improvements are the use of: (1) correct vertex function
for $\gamma N\rightarrow\Delta$, as given in Eq.
(\ref{Vertex-D3-N-to-D}); (2)  realistic form factors for the
$\Delta$; and (3) a realistic set of values for the magnetic dipole,
electric quadrupole, and Coulomb quadrupole excitation strength
for the $N\rightarrow\Delta$ transition as recently extracted from
experiments.  We demonstrate by explicit calculations that each of
these three improvements incurs considerable change in predictions
for the reduced cross sections. We then proceed to calculate, with
the three improvements implemented together, the contributions of
TPE arising from both nucleon and $\Delta$ intermediate states, to
all unpolarized and polarized observables which have been measured
or proposed in order to unravel possible causes underlying the
discrepancy in the determination of $R$. They include the
unpolarized cross sections, extracted value of $R$ in LT method,
ratio $R^\pm$ between the positron-proton and electron-proton cross
sections, beam-normal and target-normal single spin asymmetries, and
the transverse and longitudinal polarizations of the recoil proton,
$P_t$ and $P_l$, and their ratio $R_{PT}=-\mu_p
\sqrt{\tau(1+\epsilon)/2\epsilon}P_t/P_l$.

For the TPE correction to the unpolarized cross sections associated with  the $\Delta$
 intermediate states (TPE-$\Delta$), our results for $\delta_\Delta$ show a peculiar behavior of rapidly rising or decreasing as $\epsilon \rightarrow
 1$. We argue that our hadronic model is not expected to be reliable at large energies and should be restricted  for $W \leq 3-4$ GeV. For  $W = 3.5$ GeV it gives $\epsilon <0.904$ at $Q^2=3$ GeV$^2$. We hence limit the comparisons of our predictions with  the experimental data at low $Q^2$ located in this region throughout this study. Moreover, $\delta_\Delta$'s we obtain are substantially larger
 than the DR results of \cite{BK12,BK14}. We speculate that the difference most likely arises from asymptotic behavior of the TPE-$\Delta$ amplitude at $s\rightarrow \infty$ which is assumed to vanish
 in the DR calculations. Whether the assumption that TPE amplitude would vanish at $\epsilon\rightarrow 1$, i.e., $s\rightarrow \infty$ limit, at soft hadronic scale should be an interesting question to pursue further.

 We find that  the combined TPE effects of
 TPE-N and  TPE-$\Delta$  as prescribed in our hadronic model,
 can give a reasonable explanation of the the data measured in 1994 by Andivahis {\it et al.}
 \cite{Andivahis94}. The values of $R$ extracted from this data set, with TPE effects taken into account,
 are also  close to the PT values. However, this sweet agreement turns sour when the recent high precision
 super-Rosenbluth data measured at Jlab as well the 1994 data of \cite{Walker94} are analyzed, with TPE effects
 accounting for less than $50\%$ of the discrepancy between LT and PT values.

 The values of the ratio $R^\pm$ between $e^\pm p$ scatterings predicted by our model, appear to be
 in reasonable agreement with the preliminary results from VEPP-3 \cite{VEPP3-2012,VEPP3-2014}, except for one data point.
 This might indicate that the real part of the amplitude prescribed by our hadronic model is
 not unsatisfactory, at least in the low $Q^2$ region.
 Better understanding would come only after both VEPP-3 and CLAS \cite{Weinstein13} finish their
 analyses as well as more data at higher $Q^2$ region. However, our predictions show considerable variance
 with the data of Moteabbed {\it et al.} \cite{Moteabbed13} which were measured at large $\epsilon$.
 The data seem to remain finite as $\epsilon\rightarrow 1$ which contradicts the general expectation that
 TPE corrections to $\sigma_R$ would approach zero. Whether this is an artifact of the large uncertainty in the
 beam luminosity in the experiment of \cite{Moteabbed13}, or it can be used to support our results that
 effects of TPE-$\Delta$ for $\sigma_R$ show an anomalous behaviour there is real, should be studied further.

 For the angular distributions of the beam-normal   spin asymmetry $B_n$, our predictions are too large at
 $\theta_{cm}\sim 60^\circ$,
 where there are only two data points available in the energy region in which our model, with only N and $\Delta$
 intermediate states included, is expected to be applicable. However, we are encouraged to see that
 our predictions for the variation of $B_n \,\,vs. \,\,E_e$, appear to be in satisfactory agreement with data at
 larger angles $\theta_{cm}\sim 120-150^\circ$, except one data point at $E_e \sim 0.7$ GeV and
 $\theta_{cm}\sim 120^\circ$. For the target-normal spin asymmetry
 $A_n$, no data are available for comparison. Our results for the
 angular distributions at lower energies agree, in general, with
 results of \cite{Pasquini04}. However, considerable differences,
 not only in magnitude but also in shape, appear as energy
 increases. It could arise from the treatment of $\Delta$ width and
 the contributions of higher nucleon resonances.

 For the polarization observables $P_t, P_l$ and the ratio
 $R_{PT}$, we find that the contribution of TPE-$\Delta$ is smaller
 than that of TPE-N. Taken together, our hadronic model fails to explain the recent measurement of
 $P_l/P_l^{1\gamma}$ by GEP$2\gamma$ at Jlab \cite{Meziane11} for
 $\epsilon > 0.6$. Besides, the addition of the effect of TPE-$\Delta$
 appears to slightly shift upward by about $2\%$, the reasonable description of the data
 on $R_{PT} \,\,vs. \,\,\epsilon$ by TPE-N alone.

 Several questions have arisen from our study. The first one
 concerns the large difference in the extracted values of $R$ from
 data94 of \cite{Andivahis94} and data06 of \cite{Qattan06}, both
 before and after the TPE corrections are implemented. We have
 little clue about this and experimentalists might be of much help
 in this regard. Taken together the encouraging results from
 analyzing data94 and the reasonable agreement found between our
 predictions for $R^\pm$ and the preliminary data from VEPP-3, one
 is tempted to say that the real part of the amplitude as prescribed
 from our model might not be very far from realistic, at least in the low $Q^2$ region,
 especially if the further analyses from VEPP-3 and CLAS will confirm our
 predictions. Our model descriptions of the polarization data of
 beam-normal asymmetry $B_n$
 and recoil proton polarizations $P_l$ and $R_{PT}$ range from
 good to poor. The disagreement between our predictions and some of
 the polarization data raise intriguing challenge to our model.
 Since the polarization observables like single spin asymmetries are closely connected with the imaginary part of the TPE amplitude, one could immediately ask whether the recipe we follow to account for the effect of the $\Delta$ width is reasonable. In addition, theoretical
 questions like the off-shell effects of the $\Delta$ and the
 contributions of the $\pi N$ continuum and higher nucleon resonances which have been studied in \cite{Kondratyuk07,BK14} also deserve more careful study. Other possible TPE mechanisms, like the t-channel meson exchange processes as suggested in \cite{chen14}, should be explored further as well.

\vspace{1.0Cm} \centerline{\bf Acknowledgments}

We thank J.
Arrington, P. G. Blunden, C. W. Kao, and B. Pasquini for helpful
communications and discussions.  S.N.Y. would like to dedicate this
work to the memory of John A. Tjon. This work is supported in part
by the National Natural Science Foundations of China under Grant No.
11375044, the Fundamental Research Funds for the Central Universities under Grant No. 2242014R30012 for H.Q.Z. and the National Science Council of the Republic of China (Taiwan) for S.N.Y. under grant No. NSC101-2112-M002-025.
H.Q.Z. would also like to gratefully acknowledge the support of
the National Center for Theoretical Science (North) of the National
Science Council of the Republic of China for his visit in the summer
of 2012.

\end{document}